\documentclass[%
 reprint,
 amsmath,amssymb,
 aps,nofootinbib
]{revtex4-1}

\usepackage[subnum]{cases}
\usepackage{blkarray}
\usepackage{verbatim} % comentarios

\usepackage{graphicx}% Include figure files
\usepackage{dcolumn}% Align table columns on decimal point
\usepackage{bm}% bold math
\begin{document}

%\preprint{APS/123-QED}

\title{CP symmetry violation in the scalar sector of 331 models}% Force line breaks with \\

\author{Camilo A. Rojas}
\email{carojaspac@unal.edu.co}
 %\altaffiliation[Also at ]{UAB}%Lines break automatically or can be forced with \\
\author{F. Ochoa}%
\email{faochoap@unal.edu.co}
\author{R. Martinez}
\email{remartinezm@unal.edu.co}
\affiliation{Departamento de F\'isica, Universidad Nacional de Colombia, Ciudad Universitaria\\ K. 45 No. 26-85, Bogot\'a D.C., Colombia }


\date{\today}% It is always \today, today,

\begin{abstract}
In order to understand some frameworks for CP Violation scenarios in the scalar sector, a 331 model was considered which its main property is the incorporation of a local group symmetry SU(3) in the electroweak sector. In particular, a 331 model with free parameter $\beta = 1/\sqrt{3}$ was chosen. CP Violation scenarios were obtained by introducing a discrete symmetry in the scalar triplets, which exhibit a spontaneous CP Violation frame with just one independent CP phase associated. Mass state rotations were obtained.

\begin{description}
\item[Keywords:]
CP Violation, 331 model, Complex Scalar Triplets, $Z_2$ Symmetry.
\end{description}

\end{abstract}

%\keywords{Suggested keywords}{CP symmetry, baryogenesis, 331 model, Higgs Triplet}%Use showkeys class option if keyword
\maketitle

%\tableofcontents

%===============================//============================
\section{Introduction}
%===============================//============================

The models based on the gauge symmetry group $SU(3)_c\otimes SU(3)_L \otimes U(1)_X$, called 331 models, extend the electroweak sector of the Standard Model (SM) with many interesting consequences. For example, they can explain the existence of three fermion families by considering the cancellation of chiral anomalies and the asymptotic freedom in QCD \cite{pisano1992,*Frampton_1992,buras2013B}. In addition, since the model is not universal of flavor families (the third family transform with a different representation), there arises an approach to understand the hierarchy of masses in the quark sector \cite{ochoa2005,buras2013B}. In scenarios associated with $B$ physics, we can use specific versions of 331 models to explore new physics signals, as for example, in flavor changing neutral current (FCNC) contributions at tree level induced by the new $Z'$ boson emerged from this model \cite{martinez2008,montano2013B,cogollo2012,promberger2007flavor}.\\

Another interesting aspect that can be study in the framework of 331 models, are their charge-parity symmetry (CP) properties. %exhibited by these models. %It is necessary to have a detailed understanding of the CP violation phenomena in the theoretical and experimental framework. 
 For example, one important scenario where CP violation plays a fundamental role is in the mechanism of leptogenesis \cite{WEINBERG_1979,*YOSHIMURA:1979aa,*Barr:1979ye,*NANOPOULOS:1979aa,*FRY_1980,*KOLB:1980aa,fukugita_1986,*Luty:1992un} which provide us an attractive explanation for the generation of the observed Baryon Asymmetry of the Universe (BAU). In the framework of the Electroweak Baryogenesis (EWB), the three necessary conditions proposed by Sakharov \cite{sakharov1967violation} are fulfilled in the extension of the SM with right-handed neutrinos. In fact, the Majorana masses violate the lepton number through the sphaleron processes \cite{Manton:1983nd,*Kuzmin:1985mm,fukugita_1986,*Luty:1992un,barbieri_2003,*Giudice:2004aa,*nardi_2008}, while CP violation occurs between lepton doublets and the right-handed neutrinos in the Yukawa couplings, and out-of-equilibrium neutrino decay arise \cite{fukugita_1986,*Luty:1992un,Covi:1996aa,barbieri_2003,*Giudice:2004aa,*nardi_2008}. In the SM, because of the large Higgs mass, it is not possible to obtain an out-of-equilibrium stage at the electroweak phase transition, while the CKM phase is not large enough \cite{kajantie_phase_1995}, then the resultant CP violation is too small \cite{SMCPV_93,SMCPV_quimbay_94}. Thus, feasible models of EWB need a modification of the scalar potential in order to introduce new sources of CP violation. In the 331 models, %presents the possibility of addressing the process of leptogenesis, in particular, 
 the leptonic number violation can be obtained through the contribution of new singlet neutrinos with heavy Majorana masses ($m_N \ll \upsilon_{EW}$) \cite{concha_neutrino_2003,nardi_2008}.%\footnote{The Majorana mass must be larger than the electroweak breaking scale ($m_N > \Lambda_{EW}$) in order to introduce the see-saw mechanism \cite{concha_neutrino_2003,nardi_2008}.\\

In order to incorporate CP violation, the introduction of complex couplings in the Higgs potential or complex vacuum expectation values (VEV’s) in the scalar fields can be proposed to generate explicit or spontaneous CP violation. In multi-Higgs models, as the popular two-Higgs-doublet model (2HDM), CP violation in the scalar sector can be controlled through discrete symmetries. Usually, various CP-violation terms appears in the Higgs potential. These terms can be forbidden by requiring discrete symmetries, as for example $Z_2$, obtaining a CP-conservative model. However, by soft-breaking the discrete symmetries accepting quadratic non-invariant terms, it is possible to obtain CP-violation interactions.  

In this work, we address the CP properties of the scalar sector in the $\beta = 1/\sqrt{3}$ 331 model; specifically, we do a systematic study for spontaneous CP breaking. In contrast to the 2HDM, we control the CP-violation terms through global continuous symmetries. In this scenario, we can obtain spontaneous CP-violation by requiring: 1.) complex VEVs of the scalar sector and 2.) the breaking of the global symmetry into a discrete symmetry.    

This article is organized as follows. First, the next section is devoted to review the general properties of the CP conservative 331 models. Later, in section (\ref{CPviolation_sec}); an extension of the 331 model with CP violation in the Higgs sector, including minimal conditions for its realization is proposed. In particular, the model explores both explicitly and spontaneous CP violation. Finally, rotations into mass eigenstates are obtained in section (\ref{mass_sect}), where CP-even and -odd states are mixed.  

%Finally, we discuss some phenomenological prospects for future studies in order to test the predictions and consequences of the model.

%===============================//============================
\section{331 model}
%===============================//============================

A 331 model is a group extension of the SM group to the $SU(3)_C \otimes SU(3)_L \otimes U(1)_X$ gauge symmetries. One consequence of this extension, is that the 331 models have an extended Higgs sector, and a two-stage Spontaneous Symmetry Breaking (SSB) is required, where the first transition occurs at a higher scale than the electroweak breaking \cite{buras2013B}. Also, they introduce new particles in the fermionic sector, which may produce new physics. In addition, the extension of the SM gauge group from $SU(2)_L$ to $SU(3)_L$ implies the existence of five new gauge bosons.\\

The $SU(3)_L$ group must satisfy the Lie algebra, thus the generators are proportional to the Gell Mann matrices; $G_{\alpha} = \frac{1}{2}\lambda_{\alpha}$ with $\alpha = 1, ..., 8$. The electric charge operator is defined as the following combination of the diagonal generators:

\begin{equation} \label{eqhj}
\hat{Q} = \alpha G_3 + \beta G_8 + X  
,\end{equation}
where we introduced the new quantum number ($X$), proportional to the identity matrix. By requiring the Gell-Mann Nijishima relation; $\hat{Q} = G_3 + \hat{Y} $, where $G_3$ is an extension of the third Pauli matrix $\sigma_3$, we may choose $\alpha = 1$ and $\hat{Y} = \beta G_8 + X$ for the hipercharge, obtaining $\beta$ as a free parameter of the model. The phenomenology of 331 models depends on the value assigned to the $\beta$ parameter \cite{diaz2004,diaz2005,ochoa2005}. In particular, in order to avoid exotic charges (non-SM electric charges) this parameter must take the values $\beta = \pm \frac{1}{\sqrt{3}}$.

%\begin{equation} \beta = \frac{1}{\sqrt{3}} .\end{equation}

Finally, to be consistent with the SM phenomenology, the new particles must acquire masses at a very high energy. This can be achieved in a natural way by introducing a two-stage SSB scheme at different energy scales, where the new particles become massive in the first stage at a high energy scale above the electroweak scale. 

\subsection{Particles Spectrum}
%\subsection{Fermionic Spectrum}\label{ferm_spectrum}

In 331 models, the left-handed leptons form triplets or antitriplets, and the right-handed leptons have singlet representations, similar to the SM, while in the quark sector, two left-handed generations comes in triplets while the other family has an anti-triplet representation in order to cancel the chiral anomalies. Thus, in the quark sector, the model is naturally nonuniversal of families, which provide a hint to understand the hierarchy mass problem exhibited by the heaviest family and the other ones. The above structure is summarize as follows:

\begin{eqnarray}
\nonumber Q_L^{(a)} = \begin{pmatrix}u^{(a)}\\d^{(a)} \\ J^{(a)} \end{pmatrix}_L \;\;\;\;&\sim&\;\;\;\; (3,3,0)\\
\nonumber Q_L^{(3)} = \begin{pmatrix}b\\ -t \\ J^{(3)} \end{pmatrix}_L \;\;\;\;&\sim&\;\;\;\; (3,3^*,1/3)\\
\nonumber \ell_L^{(i)} = \begin{pmatrix} e^{(i)} \\ -\nu^{(i)} \\ N^{(i)} \end{pmatrix}_L \;\;\;\;&\sim&\;\;\;\; (1,3^*,-1/3)
\end{eqnarray}
%------ aca realmente los numeros cuanticos del anti triplete me dan de signo contrario aplicando el operador carga de signo contrario sobre el antitriplete pero no entiendo porque se deben cambiar los signos en la segunda componente y del numero cuantico X sin que afecte los calculos de la carga ??? eso no me da !!! ver referencia: a buras and t anatomy of quarks 
\begin{eqnarray}
\nonumber u_R^{(i)}, J_R^{(3)}  \;\;\;\;\;\; &\sim&\;\;\;\; (3,1,2/3)\\
%\nonumber J_R^{(3)} \;\;\;\;\;\; &\sim&\;\;\;\; (3,1,2/3)\\
\nonumber d_R^{(i)}, J_R^{(a)} \;\;\;\;\;\; &\sim&\;\;\;\; (3,1,-1/3)\\
%\nonumber J_R^{(a)} \;\;\;\;\;\; &\sim&\;\;\;\; (3,1,-1/3)\\
\nonumber e_R^{(i)} \;\;\;\;\;\; &\sim&\;\;\;\; (1,1,-1)\\
\nonumber \nu_R^{(i)}\;,\;N_R^{(i)} \;\;\;\;\;\; &\sim&\;\;\;\; (1,1,0).
\end{eqnarray}
% cabe recordar que se tomo el espacio dual sobre los leptones pero es una opcion arbitraria para que los nuevos leptones sean tipo neutrinos (sin interaccion electromagnetica)----------  
In the fermionic spectrum shown above, we indicate the corresponding ($SU(3)_C,SU(3)_L,U(1)_X$) representations, where $J^{(i)}$ are new quarks, and $N^{(i)}$ are new neutral leptons. The spectrum also includes right-handed neutrinos which give us a better framework to address the phenomenology of neutrinos. The superscript convention is: $a =1,2$ for the first two families and $i=1,2,3$ run over all the three families.\\

%\subsection{Bosonic Spectrum}

In the bosonic sector, the covariant derivative over triplets is :
\begin{equation}
D_{\mu} =\partial_{\mu} - igW^{\alpha}_{\mu}G_{\alpha} - ig_XXB_{\mu} 
\label{derivative_331}
,\end{equation}
where we must introduce 9 vector fields: eight fields $W^{a}_{\mu}$ associated with each $SU(3)$ generator and one field $B_{\mu}$, associated to the $U(1)$ group. We obtain the following representation for the gauge fields:

\begin{equation} W_{\mu} = \frac{1}{2} \begin{pmatrix} W^3 + \frac{1}{\sqrt{3}}W^8 & \sqrt{2} W_{\mu}^{+} & \sqrt{2} V_{\mu}^{+} \\ \sqrt{2} W_{\mu}^{-}  & -W^3 + \frac{1}{\sqrt{3}}W^8 & \sqrt{2} V_{\mu}^{0} \\ \sqrt{2} V_{\mu}^{-} & \bar{V}_{\mu}^{0} & -\frac{2}{\sqrt{3}}W^8 \end{pmatrix}
.\end{equation}
%where the following definitions have been introduced:
%\begin{equation*} U_{\mu}^{\pm}=\frac{1}{\sqrt{2}}(W^1_{\mu} \mp iW^2_{\mu}) \;\;\; , \;\;\; V_{\mu}^{\pm}=\frac{1}{\sqrt{2}}(W^4_{\mu} \mp iW^5_{\mu}) ,\end{equation*}
%\begin{equation*} V_{\mu}^{0}=\frac{1}{\sqrt{2}}(W^6_{\mu} - iW^7_{\mu}) \;\;\; , \;\;\; \bar{V_{\mu}^{0}}=\frac{1}{\sqrt{2}}(W^6_{\mu} + iW^7_{\mu}) .\end{equation*}
%In addition, for the $U(1)$ group, we have: $\mathbb{B}_{\mu}= \mathbb{I}B_{\mu}$. 
where $W^3_{\mu}\,,\, W^8_{\mu}\,,\, B_{\mu}$ are three neutral vector bosons, which after rotations into mass eigenstates become the photon, the neutral weak boson $Z$ and a new neutral weak boson $Z'$. In addition, there arise other two neutral gauge bosons $V_{\mu}^{0}$ and $\bar{V}_{\mu}^{0}$. For the charged sector, we obtain the weak bosons $W^{\pm}$ and new charged weak bosons $V^{\pm}$. \\

On the other hand, an appropriate scalar sector must be introduced to provide the SSB mechanism that gives masses to the vector gauge bosons. In addition, the Yukawa Lagrangian which couple scalar fields with fermions, must provide masses to all fermions. In our case, the 331 group exhibit a SSB scheme that must contain at least two transitions: one that lead us from 331 to the SM gauge symmetry (321), and another one from 321 to the chromodynamics and electrodynamics symmetries (31). In order to provide masses to all fermions, we must to introduce two scalar triplets in the second transition. So, the model exhibits the following breaking scheme through their scalar Vacuum Expectation Values (VEV):  
\begin{equation*}
 SU(3)_L \otimes U(1)_X \;\; \xrightarrow[1.T.]{\;\langle\chi\rangle\;} \;\;  SU(2)_L \otimes U(1)_Y \;\; \xrightarrow[2.T.]{\;\langle\rho,\eta\rangle\;} \;\; U(1)_Q
.\end{equation*}

In the first transition (1.T.): we have five broken generators due to the VEV $\langle\chi\rangle$, where the five new gauge bosons acquire masses. In the 2.T. due to $\langle\rho,\eta\rangle$, three other generators are broken, which lead us to three massive gauge bosons and one massless gauge boson (the photon). The representation of the scalar fields are summarized in Table \ref{scalar_spect}.\\ 

%Applying charge and hypercharge operators, we can introduce the following spectrum for the scalar triplets (giving in table\ref{scalar_spect}).\\

\begin{table}[h]
\centering
\caption{Scalar Spectrum for 331 model with $\beta=\frac{1}{\sqrt{3}}$ }
\label{scalar_spect}
\begin{tabular}{c | c c c }
\hline
\hline
scalar triplet & $Q_{\Phi}$ & $Y_{\Phi}$ & $X_{\Phi}$ \\
\hline
$\chi = \begin{pmatrix} \chi^{\pm} \\ \frac{1}{\sqrt{2}} \left(\xi_{\chi_2} \pm i\zeta_{\chi_2}\right) \\ \frac{1}{\sqrt{2}} \left(\xi_{\chi_3} \pm i\zeta_{\chi_3} \right)\end{pmatrix}$ & $\begin{pmatrix} \pm1 \\ 0 \\ 0 \end{pmatrix}$ & $\begin{pmatrix} \pm\frac{1}{2} \\ \pm\frac{1}{2} \\ 0 \end{pmatrix}$ & $\frac{1}{3}$ \\ 
$\rho = \begin{pmatrix} \rho^{\pm} \\ \frac{1}{\sqrt{2}} \left( \xi_{\rho_2} \pm i\zeta_{\rho_2} \right) \\ \frac{1}{\sqrt{2}} \left( \xi_{\rho_3} \pm i\zeta_{\rho_3} \right) \end{pmatrix}$ & $\begin{pmatrix} \pm1 \\ 0 \\ 0 \end{pmatrix}$ & $\begin{pmatrix} \pm\frac{1}{2} \\ \pm\frac{1}{2} \\ 0 \end{pmatrix}$ & $\frac{1}{3}$ \\
$ \eta = \begin{pmatrix} \frac{1}{\sqrt{2}} \left(\xi_{\eta} \pm i\zeta_{\eta} \right) \\ \eta^{\mp}_2 \\ \eta^{\mp}_3 \end{pmatrix}$ & $\begin{pmatrix} 0 \\ \mp1 \\ \mp1 \end{pmatrix}$ & $\begin{pmatrix} \mp\frac{1}{2} \\ \mp\frac{1}{2} \\ \mp1 \end{pmatrix}$ & $-\frac{2}{3}$ \\
\hline  
\hline
\end{tabular}
\end{table}

In order to obtain the above SSB pattern, the most general VEV structure of each triplet is:

\begin{equation}\label{vevs_reales}
\langle\chi\rangle = \begin{pmatrix} 0 \\ 0 \\ v_{\chi} \end{pmatrix} \;\;\; , \;\;\;
\langle\rho\rangle = \begin{pmatrix} 0 \\ v_{\rho_1} \\ v_{\rho_2} \end{pmatrix} \;\;\; , \;\;\; \langle\eta\rangle = \begin{pmatrix} v_{\eta} \\ 0 \\ 0 \end{pmatrix} 
.\end{equation}

Since each field $\rho$ and $\eta$ does not fit VEVs separately in the first and second components simultaneously, it is necessary to take both scalar fields in the second transition in order to ensure that all fermions become massive after the SSB. %In fact, we \\

As we can see in Table \ref{scalar_spect}, the scalar triplets that breaks the symmetry in the second transition, $\rho$ and $\eta$, have different quantum number on $X$, defining an unique scalar basis for CP studies, in contrast to models as in 2HDM which exhibits two identical scalar multiplets, and any combination between them define one possible basis that respect the gauge symmetry. Thus, we do not need to propose a general CP transformation, as occurs with 2HDM \cite{ecker1987GCP,branco2011}.

\subsection{Higgs and Yukawa Lagrangian}
%\subsection{Higgs Lagrangian}

%For the kinetic terms, we can set the sum of three terms corresponding to the three triplets. For our purpose, we focus on the Higgs potential, which has not a unique notation, so was chosen the most general which is convenient for CP studies. 
%In order to construct the Higgs potential terms, those must be hermitian, renormalizable and gauge invariant in $SU(3)_L \otimes U(1)_X$. 

The most general hermitian, renormalizable and gauge invariant Higgs potential, considering $\beta=\frac{1}{\sqrt{3}}$, is:

\begin{eqnarray}
\nonumber V_H = && \mu^2_{1} \chi^{\dagger}\chi + \mu^2_{2} \rho^{\dagger}\rho + \mu^2_3 \eta^{\dagger}\eta + ( \mu^2_4 \chi^{\dagger}\rho + h.c. )\\
\nonumber &+& ( f \epsilon^{ijk}\eta_i\rho_j\chi_k +h.c. ) \\
\nonumber &+& l_1(\chi^{\dagger}\chi)^2 + l_2 (\rho^{\dagger}\rho)^2 + l_3 (\eta^{\dagger}\eta)^2\\
\nonumber &+& l_4 \chi^{\dagger}\chi \rho^{\dagger}\rho  + l_5 \chi^{\dagger}\chi \eta^{\dagger}\eta + l_6 \rho^{\dagger}\rho \eta^{\dagger}\eta + l_7 \chi^{\dagger}\eta \eta^{\dagger}\chi \\
\nonumber &+& l_8 \chi^{\dagger}\rho \rho^{\dagger}\chi + l_9 \eta^{\dagger}\rho \rho^{\dagger}\eta + (l_{10} \chi^{\dagger}\rho \chi^{\dagger}\rho + h.c.)\\ 
\nonumber &+& (l_{11} \rho^{\dagger}\rho \rho^{\dagger}\chi + h.c.) + (l_{12} \eta^{\dagger}\eta \chi^{\dagger}\rho + h.c.) \\
\label{whole_potential} &+& (l_{13} \chi^{\dagger}\chi \chi^{\dagger}\rho + h.c.) + (l_{14} \eta^{\dagger}\chi \rho^{\dagger}\eta + h.c.),
\end{eqnarray}
%\subsection{Yukawa Lagrangian}
while the most general Yukawa Lagrangian that couple left- and right-handed quarks to Higgs fields has the following structure:
%\subsubsection{Quarks}

\begin{eqnarray}
\nonumber \mathcal{L}_{Y}^{ _Q} &=& \bar{Q}_L^{(a)}\left( \Gamma^{(_d,\rho)}_{a,i}\,\rho d^{(i)}_R + \Gamma^{(_u,\eta)}_{a,i}\, \eta u^{(i)}_R + \Gamma^{(_J,\chi)}_{a,a'}\, \chi J^{(a')}_R \right)\\
\nonumber &+& \bar{Q}^{(3)}_L\left( \Gamma^{(_d,\eta)}_{3,i}\eta^{\ast} d^{(i)}_R + \Gamma^{(_u,\rho)}_{3,i}\rho^{\ast} u^{(i)}_R + \Gamma^{(_J,\chi)}_{3,3}\chi^{\ast} J^{(3)}_R \right)\\
\nonumber &+& \Gamma^{(_J,\rho)}_{a,a'}\bar{Q}^{(a)}_L \rho J^{(a')}_R + \Gamma^{(_J,\rho)}_{3,3} \bar{Q}^{(3)}_L \rho^{\ast} J^{(3)}_R\\
\nonumber &+& \Gamma^{(_J,\eta)}_{3,a'} \bar{Q}^{(3)}_L \eta^{\ast} J^{(a')}_R + \Gamma^{(_J,\eta)}_{a,3} \bar{Q}^{(a)}_L \eta J^{(3)}_R \\
\label{general_yukawa_quarks} &+& \Gamma^{(_d,\chi)}_{a,i} \bar{Q}^{(a)}_L \chi d^{(i)}_R + \Gamma^{(_u,\chi)}_{3,i} \bar{Q}^{(3)}_L \chi^{\ast} u^{(i)}_R + h.c.
,\end{eqnarray}
%where indices convention is maintained. 

We can note in the above Lagrangian, that the new quarks $J$ obtain masses from the coupling to the scalar triplet $\chi$ in the first transition. Also, there exists mixing terms with the light sector through the triplets $\rho$ and $\eta$ at the scale of the second transition. Although the gauge invariance does not forbid those terms, we can restrict these mixing terms through extra global symmetries.\\  

%\subsubsection{Leptons}
Analogous to the quark sector, for the leptonic part in the Lagrangian we have the following terms:

\begin{eqnarray}
\nonumber \mathcal{L}_{Y}^{\ell} = && \overline{\ell_L^{(i)}}\left( \Gamma^{(\nu,\rho)}_{i,j}\,\rho^{\ast} \nu^{(j)}_R + \Gamma^{(e,\eta)}_{i,j}\, \eta^{\ast} e^{(j)}_R + \Gamma^{(_L,\chi)}_{i,j}\, \chi^{\ast} N^{(j)}_R \right)\\ 
\nonumber &+& \overline{\ell^{(i)}_L}\left( \Gamma^{(_L,\rho)}_{i,j}\rho^{\ast} N^{(j)}_R + \Gamma^{(\nu,\chi)}_{i,j}\chi^{\ast} \nu^{(j)}_R \right) + h.c.\\
\label{general_yukawa_leptons} &-& \frac{1}{2}\left(m_{\nu}\overline{\nu_R^{c}}\nu_R + m_{N}\overline{N_R^c}N_R\right)_{(i)}
,\end{eqnarray}

\noindent where couplings with right-handed neutrinos and Majorana terms are introduced.

%===============================//============================
\section{CP Violation}\label{CPviolation_sec}
%===============================//============================

CP violation in the scalar sector can be introduced in two forms: explicit or spontaneous. The first form occurs when some terms in the scalar Lagrangian are non-invariant under the CP transformation. In the second one, the Lagrangian of the model conserves CP, but the vacuum of the scalar sector is not CP-invariant, so that the theory breaks CP together with the SSB \cite{haber2012group}. In the above section, if we consider that all the Lagrangian parameters and VEVs are real, then the model does not break CP neither explicitly nor spontaneously. Now, we are going to consider the situation when couplings and VEVs are complex. In the literature, some particular CP-violation scenarios in these models have already been considered before \cite{promberger2007flavor,doff2006spontaneous,montero1999soft}. Here, we undertake a systematic study of the conditions for CP violation in the model with $\beta =1/\sqrt{3}$. For that, we first note that only seven of the nineteen couplings of the Higgs potential in equation (\ref{whole_potential}) can be complex. They are: one quadratic coupling ($\mu _4^2$), one cubic coupling  ($f$), and five quartic couplings ($l_{10,11,12,13,14}$). Thus, we may introduce up to 7 phases in the potential. In addition, the SSB mechanism allows four VEVs as shown in equation (\ref{vevs_reales}), which may introduce 4 more complex phases as:

\begin{equation}\label{VEV_complex}
\langle\chi\rangle = \begin{pmatrix} 0 \\ 0 \\ |v_{\chi }|e^{i\alpha_{\chi}} \end{pmatrix},
\langle\rho\rangle = \begin{pmatrix} 0 \\ |v_{\rho_1}|e^{i\alpha_{\rho_1}} \\ |v_{\rho_2}|e^{i\alpha_{\rho_2}} \end{pmatrix},
\langle\eta\rangle = \begin{pmatrix} |v_{\eta}|e^{i\alpha_{\eta}} \\ 0 \\ 0 \end{pmatrix}
.\end{equation}
However, because of the rephasing invariance of the potential, most of the above phases can be reabsorbed, and not all of them will lead us to interactions with CP violation. Furthermore, because of the gauge symmetry, we may induce appropriate transformations to eliminate two VEVs, as shown below.       \\

%In the model considered, we have two triplets which break local gauge group to $U(1)_Q$ in the second transition and these triplets have different quantum numbers, so we have a unique basis to analyze explicit or spontaneous CP violation. In this chapter we consider the same structure for 331 model as in the previous chapter with the difference that we are going to introduce complex \emph{VEV's} and some global symmetries.

%\subsection{Complex VEV's}

%In the previous chapter, real vacuum expectation values were introduced in (Eq. \ref{vevs_reales}). Now, we consider the most general complex \emph{VEV's} structure as:

%\begin{equation}\label{VEV_complex}
%\langle\chi\rangle = \begin{pmatrix} 0 \\ 0 \\ |u|e^{i\alpha_{\chi}} \end{pmatrix},
%\langle\rho\rangle = \begin{pmatrix} 0 \\ |\nu_1|e^{i\alpha_{\rho_1}} \\ |\nu_2|e^{i\alpha_{\rho_2}} \end{pmatrix},
%\langle\eta\rangle = \begin{pmatrix} |\omega|e^{i\alpha_{\eta}} \\ 0 \\ 0 \end{pmatrix}
%.\end{equation}

%Taking into account the previous expressions for the triplets, we have initially 4 phases associated with CP violation phenomenology, however, these phases may or not be physical phases. So, we must check how the triplets transform by rotations of $SU(3)_L$ and $U(1)_X$ local groups and also it is necessary to find the stationary solutions for the Higgs potential.

\subsubsection{$SU(3)_L$ Rotation}

The Higgs potential must be invariant under transformations of the $SU(3)_L$ group. First, we define the general transformation over the scalar fields as:

\begin{equation}
\mathrm{SU(3)_L}: \;\;\; \phi_a \rightarrow \phi'_a = e^{i\vec{\Omega}_a \cdot \vec{T}}\phi_a\
,\end{equation}
where $\vec{T}$ are the group generators ot the $SU(3)_L$ group and $\vec{\Omega}_a$ are the transformation parameters for each field $a=\chi, \rho$ and $\eta$. Applying the transformation on the Higgs potential and by invoking the $SU(3)_L$ invariance ($V'_H=V_H$), we obtain the following conditions between the parameters:

\begin{equation}\label{angles_rot_su3}
\vec{\Omega}_{\chi}=\vec{\Omega}_{\rho} \;\;\;\; , \;\;\;\; \vec{\Omega}_{\eta}=-2\vec{\Omega}_{\rho}\;.
\end{equation}

In order to study how the scalar triplets transform, we use the spectral theorem for general $SU(N)$ groups to expand the exponential as \cite{weigert1997baker}:

%need the operator in matricial form but we can not use the Euler's formula, due to $SU(3)_L$ generators do not satisfy $(\hat{n}_a \cdot \vec{T})^2=\mathbb{I}$, therefore we can use the spectral theorem and the BCH relation for unitary special groups ($SU(N)$) \cite{weigert1997baker}. In that case, the transformation can be represented as:

\begin{equation}
 e^{i\mathbb{M}}=\sum_{n=0}^{N-1}\mathrm{f}_n(\mathbb{M})\mathbb{M}^n
\label{exp_pot}
,\end{equation}
with $\mathbb{M}$ an hermitian matrix, while the coefficients are:
% which corresponds to $\mathbb{M}=\vec{\Omega}_a \cdot \vec{T}$ in our case (for $N=3$), and the coefficients are:

\begin{eqnarray}
\mathrm{f}_n(\mathbb{M})=\sum _{j=1}^{N}P_{jn}e^{im_j},
\end{eqnarray}
where $m_j$ are the eigenvalues of the matrix $\mathbb{M}$, and $P_{jn}$ are coefficients obtained from:

\begin{eqnarray}
\prod _{i\neq j} \frac{\mathbb{M}-m_i}{m_j-m_i}=\sum _{n=0}^{N-1}P_{jn}\mathbb{M}^n.
\label{productoria}
\end{eqnarray}

In our case, for $SU(3)$, we have that $\mathbb{M}=\vec{\Omega}_a \cdot \vec{T}$, and $N=3$. Upon identifying $\vec{\Omega}_a=\alpha _a\hat{n}$ and $\hat{\mathbb{M}}=\hat{n} \cdot \vec{T}$ the following expansion can be obtained \cite{curtright2015elementary}:%By applying the Cayley-Hamilton theorem and the spectral decomposition
% The second equality in Eq. \ref{exp_pot} is obtained after using the Cayley-Hamilton theorem; $M^3=I \text{det}(M) + \frac{1}{2}M\text{tr}(M^2)$. Applying the spectral descomposition for the special unitary group $SU(3)$ and taking out the angle leaving the matrix with unitary values; $\mathbb{M}'=\Omega (\hat{n}_a \cdot \vec{T}) = \Omega \mathbb{M}$, we can apply the general formula for the $\mathrm{f}_n$ coefficients \cite{curtright2015elementary}, which are given by:

\begin{eqnarray}
\nonumber e^{i\vec{\Omega}_a \cdot \vec{T}}=&&\sum_{k=0}^2  \left[ \hat{\mathbb{M}}^2 + \frac{2}{\sqrt{3}}\sin(\phi + 2\pi k/3)\hat{\mathbb{M}}\right. \\
&&-\left. \frac{1}{3}(1+2\cos(2\phi + 4\pi k/3))\mathbb{I} \right]\mathcal{F},
\end{eqnarray}
where:
\[\mathcal{F}=\frac{e^{\frac{2}{\sqrt{3}}i\alpha _a\sin(\phi + 2\pi k/3)}}{1-2\cos(2\phi + 4\pi k/3)} ,\] 
%and matrix $\mathbb{M}$ is related with phase $\phi$ through: 
\[\phi=-\frac{1}{3}\arcsin{\left(\frac{3\sqrt{3}}{2}\text{det}(\mathbb{M})\right)}\;.\]

Making the explicit calculations, as shown in the Appendix \ref{mat_trans_su3}, the following transformation matrix is obtained:
\begin{equation}\label{mat_rot_repam}
e^{i\vec{\Omega} \cdot \vec{T}}\equiv\mathbb{T}=
\begin{pmatrix} T_1 & m_1+n_1 & m_2 + n_2 \\ m_1 - n_1 & T_2 & m_3 + n_3 \\ m_2 - n_2 & m_3 - n_3 & T_3 \end{pmatrix}
,\end{equation}
where the parameters $m_{\{1,2,3\}}$, $n_{\{1,2,3\}}$ and $T_{\{1,2,3\}}$ are defined in Equation (\ref{app:reparam}) from the appendix \ref{mat_trans_su3}.

Through the above matrix representation of a general SU(3) transformation, we can rotate the scalar triplets in (\ref{VEV_complex}) in order to reduce: \textit{1.)} the number of complex phases of the VEV's from 4 to 3, and \textit{2.)} the number of VEVs in $\langle\rho\rangle$ from 2 to 1. In particular, we choose to rotate this triplet so that one VEV remains in the second component in order to ensure that all fermions acquire masses. As shown in the Appendix \ref{mat_trans_su3}, the matrix that lead us to the reduction described above is: \\

\begin{equation}
\mathbb{T}=
\begin{pmatrix} 
\omega'/v_{\eta}& 0 & 0 \\ 
0 & v'/v_{\rho_1} & 0 
\\ 0 & -(v_{\rho_2}u')/(v_{\rho_1}v_{\chi}) & u'/v_{\chi}
\end{pmatrix}, 
\end{equation}
%\noindent Applying $\mathbb{T}$ to each scalar field in vacuum state, and taking into account the relation between the rotation angles (Eq. \ref{angles_rot_su3}), as shown in the appendix (\ref{mat_trans_su3}), we obtain the following restriction over the rotation $SU(3)_L$ matrix ($\mathbb{T}$), which eliminates one VEV in $\langle\rho\rangle$ and leaves invariant $\langle\chi\rangle$ and $\langle\eta\rangle$:
%\begin{equation}
%\resizebox{.3 \textwidth}{!}{
%$
%\mathbb{T}=
%\begin{pmatrix} \frac{m_3v_{\rho_1}}{v_{\rho_2}} & m_1 & -\frac{m_1v_{\rho_1}}{v_{\rho_2}} \\ 0 & \frac{v'}{v_{\rho_1}} & 0 \\ 0 & m_3 & -\frac{m_3v_{\rho_1}}{v_{\rho_2}} \end{pmatrix}
%$}
%,\end{equation}
obtaining a new basis with the following complex VEV structure:
%obThus, in the vacuum state the scalar triplets rotate to the new basis, in which their components are in general complex values:

\begin{equation}\label{VEV_rotados_su3}
\langle\chi\rangle = \begin{pmatrix} 0 \\ 0 \\ |u'|e^{i\alpha_{\chi}} \end{pmatrix} \; , \; 
\langle\rho\rangle = \begin{pmatrix} 0 \\ |\upsilon '|e^{i\alpha_{\rho}} \\ 0 \end{pmatrix} \; , \;
\langle\eta\rangle = \begin{pmatrix} |\omega '|e^{i\alpha_{\eta}} \\ 0 \\ 0 \end{pmatrix}.
\end{equation}
%where the new VEV's are related to the old one as:

%\begin{eqnarray*}
%u'&=&\left(\frac{\upsilon_2}{\upsilon_1 m_3}\right)^2 u\\
%\upsilon'&=&\left(\frac{\upsilon_2}{m_3}\right)^2\frac{1}{\upsilon_1}\\
%\omega'&=&2 \left(\frac{\upsilon_1 m_3}{\upsilon_2}\right) \omega
%.\end{eqnarray*}

Using this basis, three complex phases remains in the vacuum state. However, further reduction can be obtained by applying an $U(1)_X$ phase rotation, as shown below.

\subsubsection{$U(1)_X$ Rotation}

The transformation associated to $U(1)_X$ group, which operates over the scalar fields $\phi$ can be defined as:

\begin{equation}
\mathrm{U(1)_X}: \;\;\; \phi_a \rightarrow \phi'_a=e^{i\theta_a X_a}\phi_a\;,
\end{equation}
which corresponds to a phase rotation according to the value $X_a$ of each triplet. After a general $U(1)_X$ phase transformation, the terms that obtain a global phase in the scalar potential are:
% The generator for $U(1)_X$ is diagonal, so we can consider this transformation as a field rotation by a phase. It is necessary to take in mind the corresponding $X$ quantum number for each field. The terms in the potential that obtain a resultant phase are:

\begin{eqnarray}
\nonumber V_H = ... &+& ( \mu^2_{4} e^{i\frac{1}{3}(  \theta_{\chi}-  \theta_{\rho})  }\chi'^{\dagger}\rho' + h.c. ) \\ 
\nonumber &+& (f e^{i\frac{1}{3}(  \theta_{\rho}-2  \theta_{\eta}+  \theta_{\chi})  } \epsilon^{ijk}\rho'_i\eta'_j\chi'_k +h.c. ) \\
\nonumber &+& [l_{10} e^{i\frac{2}{3}(  \theta_{\chi}-  \theta_{\rho})  }\chi'^{\dagger}\rho' \chi'^{\dagger}\rho' + h.c.] \\
\nonumber &+& [ \rho'^{\dagger}\rho' (l_{11} e^{-i\frac{1}{3}(  \theta_{\chi}-  \theta_{\rho})  }\rho'^{\dagger}\chi' + h.c.)] \\
\nonumber &+& [ \eta'^{\dagger}\eta' (l_{12} e^{i\frac{1}{3}(  \theta_{\chi}-  \theta_{\rho})  }\chi'^{\dagger}\rho' + h.c.)] \\
\nonumber &+& [ \chi'^{\dagger}\chi' (l_{13} e^{i\frac{1}{3}(  \theta_{\chi}-  \theta_{\rho})  }\chi'^{\dagger}\rho' + h.c.)] \\
\nonumber &+& [l_{14} e^{-i\frac{1}{3}(  \theta_{\chi}-  \theta_{\rho})  } \eta'^{\dagger}\chi' \rho'^{\dagger}\eta' + h.c.] + ...\;.
\end{eqnarray}

\noindent From the invariance of the potential, we find that all the triplets must transform in the same amount:
%we find the following relation between the phases:

\begin{equation}
\theta_{\rho}=\theta_{\eta}=\theta_{\chi}=\theta\;.
\end{equation}

In particular, we can do a $U(1)_X$ phase rotation in order to eliminate up to one complex phase in (\ref{VEV_rotados_su3}). For example, with a rotation $\theta = -3\alpha_{\chi}$,  we can eliminate the phase associated to the scalar field $\chi$ to avoid CP violation in the first transition, obtaining the following VEV structures:
%After the $SU(3)_L$ rotation, the VEVs (eq. \ref{VEV_rotados_su3}) are:

%\begin{equation}\label{VEV_rotados_complex}
%\langle\chi\rangle = \begin{pmatrix} 0 \\ 0 \\ |u|e^{i\alpha_{\chi}} \end{pmatrix} \; , \; 
%\langle\rho\rangle = \begin{pmatrix} 0 \\ |\nu|e^{i\alpha_{\rho}} \\ 0 \end{pmatrix} \; , \;
%\langle\eta\rangle = \begin{pmatrix} |\omega|e^{i\alpha_{\eta}} \\ 0 \\ 0 \end{pmatrix}.
%\end{equation}

\begin{equation}\label{VEV_rotados_complex_2}
\langle\chi\rangle = \begin{pmatrix} 0 \\ 0 \\ |u'| \end{pmatrix} \; , \; 
\langle\rho\rangle = \begin{pmatrix} 0 \\ |\upsilon '|e^{i\delta_{\rho}} \\ 0 \end{pmatrix} \; , \;
\langle\eta\rangle = \begin{pmatrix} |\omega '|e^{i\delta_{\eta}} \\ 0 \\ 0 \end{pmatrix}
,\end{equation}
where the new phases are defined as:
\begin{equation}
\delta_{\rho}=\alpha_{\rho}-\alpha_{\chi} \;\;\;,\;\;\; \delta_{\eta}=\alpha_{\eta}+2\alpha_{\chi}\;,
\end{equation}
or, in cartesian form:

%For our purpose, we are going to consider in the next chapter the cartesian form for the above scalar triplets in the vacuum state, it is:

\begin{equation}\label{VEVs_3}
\langle\chi\rangle = \begin{pmatrix} 0 \\ 0 \\ u \end{pmatrix} \;,\;
\langle\rho\rangle = \begin{pmatrix} 0 \\ v_1 + i v_2 \\ 0 \end{pmatrix} \;,\; \langle\eta\rangle = \begin{pmatrix} w_1 + i w_2 \\ 0 \\ 0 \end{pmatrix}.  
\end{equation}

In summary, we obtain only two possible phases as candidates to produce spontaneous CP violation. However, as we will see below, from the stationary conditions of the Higgs potential, these phases are not mutually independent.

\subsection{Higgs Potential with $U(1)$ Global Symmetry}

The full Higgs potential in Equation (\ref{whole_potential}) can be separated into two parts: the first one with real couplings, and the other one with complex coupling constants, $V_H=V_{\mathbb{R}} + V_{\mathbb{C}}$, where:
\begin{eqnarray}
\nonumber V_{\mathbb{R}}=&\Big[& \mu^2_{1} \chi^{\dagger}\chi + \mu^2_{2} \rho^{\dagger}\rho + \mu^2_3 \eta^{\dagger}\eta + l_1(\chi^{\dagger}\chi)^2  \\
\nonumber &+& l_2 (\rho^{\dagger}\rho)^2 + l_3 (\eta^{\dagger}\eta)^2 + l_4 \chi^{\dagger}\chi \rho^{\dagger}\rho  + l_5 \chi^{\dagger}\chi \eta^{\dagger}\eta \\
\nonumber \label{real_potential} &+& l_6 \rho^{\dagger}\rho \eta^{\dagger}\eta + l_7 \chi^{\dagger}\eta \eta^{\dagger}\chi + l_8 \chi^{\dagger}\rho \rho^{\dagger}\chi + l_9 \eta^{\dagger}\rho \rho^{\dagger}\eta \; \Big]\\
\nonumber V_{\mathbb{C}}= &\Big[& \mu^2_4 \chi^{\dagger}\rho + f\epsilon^{ijk}\eta_i\rho_j\chi_k +l_{10} \chi^{\dagger}\rho \chi^{\dagger}\rho \\
\nonumber &+& l_{11} \rho^{\dagger}\rho \rho^{\dagger}\chi + l_{12} \eta^{\dagger}\eta \chi^{\dagger}\rho  \\
\label{complex_potential} &+& l_{13} \chi^{\dagger}\chi \chi^{\dagger}\rho + l_{14} \eta^{\dagger}\chi \rho^{\dagger}\eta + \textbf{h.c.} \Big]\;.
\end{eqnarray}

We can restrict the complex sector if we demand additional global symmetries. Let us assume, for example, invariance of the Higgs potential under the following $U(1)$ transformations:

\begin{equation}\label{U1_symmetry}
\chi \rightarrow \chi \;\;,\;\; \rho \rightarrow e^{i\Theta}\rho \;\;,\;\; \eta \rightarrow e^{-i\Theta}\eta \;,
\end{equation}

\noindent where $\chi$ remains invariant, while the other two scalar fields, $\rho$ and $\eta$, change by an opposite phase. Invariance under this transformation leaves us with only the f-term in the complex part of the potential:

\begin{equation}\label{potential_U1}
V_H=V_{\mathbb{R}} + \left( f \epsilon^{ijk}\eta_i\rho_j\chi_k  + h.c. \right)\;.
\end{equation}
 By using this simplified potential, we will evaluate below the CP-properties of the scalar sector.
 
\subsubsection{Explicit CP invariance}

In order to evaluate if the potential in (\ref{potential_U1}) violates CP or not explicitly, let us assume real VEVs. In this case, the only complex phase come from the cubic coupling constant $f=|f|e^{i\alpha _f}$ in the potential. However, explicit CP violation arises only if \textit{rephasing transformations of the fields that transform the complex parameters into real do not exist} \cite{Branco1999,ginzburg2005}. In our case, if we transform the three scalar triplets as
% we can define a base rotation for each scalar field, that absorb this phase. For instance, if we transform the:

\begin{eqnarray}
\label{rot_base_1} \chi &\rightarrow& \chi'=e^{i\theta_{\chi}}\chi, \\
\label{rot_base_2} \rho &\rightarrow& \rho'=e^{i\theta_{\rho}}\rho,\\
\label{rot_base_3} \eta &\rightarrow& \eta'=e^{i\theta_{\eta}}\eta.
\end{eqnarray}
then, the potential in Eq. (\ref{potential_U1}) become:

\begin{equation}\label{potential_U1_2}
V'_H=V_{\mathbb{R}} + \left( |f|e^{i\alpha_f} \epsilon^{ijk}e^{i(\theta_{\eta}+\theta_{\rho}+\theta_{\chi})}\eta'_i\rho'_j\chi'_k  + h.c. \right)\;.
\end{equation}

\noindent In particular, any phase transformation that satisfies the relation:

\begin{eqnarray}
\theta_{\eta}+\theta_{\rho}+\theta_{\chi} = -\alpha_f,
\label{real-basis}
\end{eqnarray}
will transform the above potential into a real one, i.e., there exists a real basis. Thus, the model with the global symmetry (\ref{U1_symmetry}) is explicitly CP invariant in the scalar sector. %For example, we can take:

%\[\theta_{\chi}=-\frac{1}{3}\alpha_f\;,\; \theta_{\rho}=-\frac{1}{3}\alpha_f\;,\; \theta_{\eta}=-\frac{1}{3}\alpha_f\;.\]

\subsubsection{Spontaneous CP invariance}\label{ch_spontaneus_CPc}

Knowing that the potential is explicitly CP invariant, we will evaluate if the model with complex vacuum given by (\ref{VEV_rotados_complex_2}) is also CP-invariant. %After a $SU(3)_L$ and $U(1)_X$ transformation we obtained: $\{\langle\rho\rangle,\,\langle\eta\rangle\}  \in \mathbb{C}$ and $\langle\chi\rangle \in \mathbb{R}$, as shown in equation (\ref{VEV_rotados_complex_2}).\\
After replacing the scalar triplets into the potential, evaluated at the vacuum state, and including the complex phase of the f-term, $ |f|e^{i\alpha_f}$, the complex part results with additional phases as: 

\begin{equation}\label{potential_U1_vacuum}
\resizebox{.42 \textwidth}{!}{
$
\langle V_H \rangle = \langle V_{\mathbb{R}} \rangle - |f||u'||\upsilon '||\omega '| \left(e^{-i(\delta_{\eta}+\delta_{\rho}-\alpha _f)} +e^{i(\delta_{\eta}+\delta_{\rho}-\alpha _f)} \right),
$}
\end{equation}
%where the minus sign comes from the anti-symmetrical property of the Levi-Civita tensor. 
obtaining terms with the global phase: $\delta=\delta_{\eta}+\delta_{\rho}-\alpha _f$. However, the vacuum must accomplish the minimum conditions with respect to the phases, i.e:

\begin{equation}
\frac{\partial \langle V_H \rangle }{\partial \delta_{\rho,\eta}}=0,
\end{equation}
from where we obtain the constraint:
%minimizing, we have the following result:

\begin{equation}
\label{rel_phases_pol} \alpha_f = \delta_{\eta}+\delta_{\rho}\;.
\end{equation}

\noindent So, the phases from the VEVs are not independent of the phase from the coupling constant $f$. Spontaneous CP violation arises if  \textit{rephasing transformations of the fields that transform all the complex VEVs into real, and simultaneously preserve the real basis do not exist}. In our case, as we saw above, a real basis is obtained through transformations of the form (\ref{rot_base_1}) - (\ref{rot_base_3}) with the phase relation given by (\ref{real-basis}). Under these transformations, the VEVs in (\ref{VEV_rotados_complex_2}) become:   %we have a relation between the phases, leaving one independent phase for the model (i.e. $\delta_{\eta}=\alpha_f - \delta_{\rho}$). Making a basis rotation over the scalar field, we can cancel the phases from the vacuum state, as follows:

\begin{eqnarray}\label{VEV_rotados_complex_3}
\langle\chi '\rangle &= &\begin{pmatrix} 0 \\ 0 \\ |u'|e^{-i\left(\theta _{\rho}+\theta _{\eta }+\alpha _f\right)} \end{pmatrix}, \; \nonumber \\
\nonumber \\
\langle\rho '\rangle &=& \begin{pmatrix} 0 \\ |\upsilon '|e^{i\left(\delta_{\rho}+\theta _{\rho}\right)} \\ 0 \end{pmatrix} ,
\langle\eta '\rangle = \begin{pmatrix} |\omega '|e^{i\left(\delta_{\eta}+\theta _{\eta}\right)} \\ 0 \\ 0 \end{pmatrix}
.\end{eqnarray}
so that, after considering the constraint (\ref{rel_phases_pol}), the phases of the VEVs are:

\begin{eqnarray}
\theta '_{\chi}&=&\theta _{\rho}+\theta _{\eta }+\delta_{\eta}+\delta_{\rho}, \nonumber \\
\theta ' _{\rho}&=&\delta_{\rho}+\theta _{\rho}, \nonumber \\
\theta ' _{\eta}&=&\delta_{\eta}+\theta _{\eta}.
\end{eqnarray}
Thus, real VEVs in the real basis can be obtained if all the above phases can be cancelled simultaneously. Indeed, if $\theta '_{\rho}=0$ and $\theta '_{\eta}=0$ (i.e. we rotate as $\theta _{\rho(\eta)}=-\delta _{\rho, (\eta)}$), we obtain automatically that also $\theta '_{\chi}=0$. 
\\ \\
In conclusion, the model with the global symmetry defined by (\ref{U1_symmetry}) is both explicitly and spontaneously CP invariant in the scalar sector.

%\noindent After replacing in the Higgs potential (Eq. \ref{potential_U1}), we find:
%\begin{eqnarray}
%\nonumber V'_H&=&V_{\mathbb{R}} + \left( |f|e^{i\alpha_f} \epsilon^{ijk}e^{-i(\delta_{\rho}-\delta_{\rho}+\alpha_f)}\eta'_i\rho'_j\chi'_k  + h.c. \right)\\
%\label{potential_u1_invariant2} &=& V_{\mathbb{R}} + |f|\left( \epsilon^{ijk}\eta'_i\rho'_j\chi'_k + \epsilon_{ijk}\eta'^i\rho'^j\chi'^k \right)\;.
%\end{eqnarray}

%All terms in this potential are real, so under the rephasing transformation, we eliminate simultaneously the phases from the \emph{VEV} and from the complex parameter $f$, obtaining a CP invariant potential, even if the fields originally had complex vacuum expected value.

\subsection{Breaking of Symmetry $U(1)$ $\rightarrow$ $Z_{2}$}

Let us now consider a less restricted scenario by breaking the global $U(1)$ symmetry into a discrete $Z_2$ symmetry. For example, if we limit the symmetry (\ref{U1_symmetry}) only for  $\Theta = \pi$, we obtain:

\begin{eqnarray}
 \left. Z_2\right|_{\pi}: \chi \rightarrow \chi, \;\;\; \eta \rightarrow -\eta, \;\;\; \rho \rightarrow -\rho \;,
\end{eqnarray}
so that the complex part of the Higgs potential preserves, in addition to the f-terms, the $l_{10}$-terms:  
 
%As we can see in the previous section, the $U(1)$ global symmetry leaves the model CP invariant in the scalar sector, even if we consider complex \emph{VEV} or complex parameters. However, we can introduce a symmetry soft breaking, considering $\Theta = \pi$ in the transformation equations (\ref{U1_symmetry}). This choice is equivalent to consider a $Z_2$ symmetry. Thus this transformation can be defined as:

\begin{equation}\label{higgs_potential_z2}
V_H=V_{\mathbb{R}} + \left( f \epsilon^{ijk}\eta_i\rho_j\chi_k  + l_{10} \chi^{\dagger}\rho \chi^{\dagger}\rho + h.c. \right).
\end{equation}

Below, we analyze the CP properties of this potential, as we did in the previous section. 
%Analyzing this potential under field rotations to find the invariance, we have to consider again two cases; explicit and spontaneous phases.

\subsubsection{Explicit CP invariance}

By assuming real VEVs ($\{\langle\chi\rangle,\,\langle\rho\rangle,\,\langle\eta\rangle\} \in \mathbb{R}$), we propose a general basis rotation as in equations (\ref{rot_base_1} - \ref{rot_base_3}), obtaining the following scalar potential:%so after replace the rotations in the proposed potential (Eq. \ref{higgs_potential_z2}), we have:

\begin{eqnarray}
\nonumber V_H = V_{\mathbb{R}} &+& \big( |f| e^{i\alpha_f} \epsilon^{ijk}e^{i(\theta_{\eta}+\theta_{\rho}+\theta_{\chi})}\eta'_i\rho'_j\chi'_k \\ 
\label{potential_z2_CPc_1} &+& |l_{10}|e^{i\alpha_{10}} e^{2\,i\,(\theta_{\chi}-\theta_{\rho})} \chi'^{\dagger}\rho' \chi'^{\dagger}\rho' + h.c. \big)
,\end{eqnarray}
where we include the complex phases of the $f$ and $l_{10}$ couplings. A real basis is found if we choose rotations that satisfy: % CP invariance for this potential is manifested, just if the following relations are satisfied:

\begin{eqnarray}
\theta_{\eta}+\theta_{\rho}+\theta_{\chi}&=&-\alpha_f \;,\nonumber \\
2(\theta_{\chi}-\theta_{\rho})&=&-\alpha_{10} \;,
\label{real-basis-2}
\end{eqnarray}
obtaining a real potential. Thus, the model with the $\left. Z_2\right|_{\pi}$ symmetry is explicitly CP invariant in the scalar sector.
%Again, we have the freedom to choose those phases, in a way that the above equations are satisfied, for example:

%\[\theta_{\chi}=-\frac{\alpha_f}{2}\;,\;\;\theta_{\eta}=-\frac{\alpha_{10}}{2}\;,\;\;\theta_{\rho}=\frac{\alpha_{10}-\alpha_f}{2}\;.\]
%Introducing them in the potential (\ref{potential_z2_CPc_1}), it is possible to set all the potential as real and therefore we have CP symmetry conservation:

\subsubsection{Spontaneous CP Violation}\label{SCPV_z2}

By considering general complex VEVs as in (\ref{VEV_rotados_complex_2}) into the scalar potential (\ref{higgs_potential_z2}), we find that the $l_{10}$-terms do not contribute, obtaining the same result as in equation (\ref{potential_U1_vacuum}). Thus, the phases of the VEVs are constrained by the same minimum condition found in (\ref{rel_phases_pol}), so that the relations in (\ref{real-basis-2}) become:

\begin{eqnarray}
\theta_{\eta}+\theta_{\rho}+\theta_{\chi}&=&-\delta _{\rho}-\delta _{\eta} \;,\nonumber \\
2(\theta_{\chi}-\theta_{\rho})&=&-\alpha_{10} \;.
\label{real-basis-3}
\end{eqnarray}

On the other hand, the rotations (\ref{rot_base_1} - \ref{rot_base_3}) lead us to new transformed VEVs with the following phases:

\begin{eqnarray}
\theta '_{\chi}&=&\theta _{\chi}, \nonumber \\
\theta '_{\rho}&=&\theta _{\rho}+\delta _{\rho}, \nonumber \\
\theta '_{\eta}&=&\theta _{\eta}+\delta _{\eta} 
\label{transf-phase},
\end{eqnarray}
which must accomplish both relations from equation (\ref{real-basis-3}) in order to obtain a real basis. Real VEVs are obtained if there exists some rotation that cancel all the above phases. For example, if we choose $\theta _{\chi}=0$, we obtain from (\ref{real-basis-3}) the solutions: $\theta _{\rho}=(1/2)\alpha _{10}$ and $\theta _{\eta}=-(1/2)\alpha _{10}-\delta _{\rho}-\delta _{\eta}$, which convert the phases in (\ref{transf-phase}) into:

\begin{eqnarray}
\theta '_{\chi}&=&0,  \nonumber \\
\theta '_{\rho}&=&\frac{1}{2}\alpha _{10}+\delta _{\rho}, \nonumber\\
\theta '_{\eta}&=&-\frac{1}{2}\alpha _{10}-\delta _{\rho}=-\theta '_{\rho}.
\end{eqnarray}
obtaining, at best, the constraint $\theta '_{\eta}=-\theta '_{\rho}$. Since we are assuming from the beginning a general complex potential with arbitrary phases $\alpha _{10}$ and $\delta _{\rho}$ different from zero, then we obtain VEVs with one irreducible complex phase \footnote{There is a very particular case for a model with $\alpha_{10} = -2\, \delta_{\rho}$, which cancel all the phases, and the model become CP conservative}. Thus, the model with the symmetry $\left. Z_2\right|_{\pi}$ is spontaneously CP violating.

\section{Mass states in CPV 331 model}\label{mass_sect}
%===============================//============================

Now that we have found a 331 model with a $Z_2$ symmetry which admits spontaneous CP violation in the scalar sector, we are going to analyze the mass eigenstates to obtain the physical spectrum.

\subsection{Physical Spectrum - Scalar Sector}

Table (\ref{scalar_spect}) define the scalar spectrum in the electroweak basis. We must rotate this basis in order to obtain the physical states for the scalar fields and identify the Goldstone bosons and the Higgs massive fields. For that, it is more convenient to take the complex VEVs of the scalar fields in its cartesian form, as in Eq.  (Eq. \ref{VEVs_3}), so that the minimum conditions of the Higgs potential become:

%Taking the potential with symmetry $Z_2$ (Eq. \ref{higgs_potential_z2}), and the scalar triplets in groundstate in cartesian form (Eq. \ref{VEVs_3}), we proceed to minimize the potential with respect to the vacuum expectation values, it is:

\[\frac{\partial \langle V_H \rangle}{\partial v_i}=0\;,\;\;\;\;\; v_i=v_1,v_2,w_1,w_2,u\;\;\;,\]
with $V_H $ the Higgs potential with $Z_2$ symmetry from (\ref{higgs_potential_z2}). First, after minimization, we find one condition that relates the VEVs of the $\rho $  and  $\eta $ triplets:
\begin{equation}\label{rel_fases_2}
w_2=-w_1\frac{v_2}{v_1}.
\end{equation}
Second, we obtain the following tadpoles equations:

\begin{eqnarray*}
\mu^2_1&=&-2 l_{1} u^{2} - \left( v_{1}^{2} + v_{2}^{2} \right) \left( l_{4} - f \left( \frac{w_{1}}{v_{1}} \right) + l_{5} \left(\frac{w_{1}}{v_{1}}\right)^{2} \right),\\ 
\mu^2_2&=&-l_{4} u^{2} + f \left( \frac{w_{1}}{v_{1}} \right) u - \left( v_{1}^{2} + v_{2}^{2} \right)\left( 2l_{2} + l_{6} \left(\frac{w_{1}}{v_{1}}\right)^{2} \right),\\
\mu^2_3&=&-l_{5} u^{2}+ f \left( \frac{v_{1}}{w_{1}} \right) u - \left( v_{1}^{2} + v_{2}^{2} \right)\left( l_{6} + 2l_{3} \left(\frac{w_{1}}{v_{1}}\right)^{2} \right).
\end{eqnarray*}

After replacing the above tadpoles equations into the Higgs potential, and through their second derivatives, we may obtain the mass matrices. For the charged fields, the mixing mass elements are: %it is possible to find the mass matrix using the scalar triplets which must be shifted as the sum of fields in vacuum state plus the excited states, it is: $\Phi=\langle\Phi\rangle + \hat{\Phi}$, taking $\hat{\Phi}$ in the electroweak basis (See Table \ref{scalar_spect}).\\

%Then, we can obtain the mass matrix terms, for both the neutral and charged sector. Due to the electric charge conservation, there are not mixing terms in the mass matrix, so we can put them in two separated blocks in a block diagonal matrix:

%\begin{equation}
%M^2=\begin{pmatrix}M^2_N & 0 \\ 0 & M^2_C \end{pmatrix}
%,\end{equation}
%where $M^2_N$ is a matrix of order 10, for the neutral sector, and the matrix $M^2_C$ is a matrix of order 4 with charged fields. 

%\subsubsection{Charged Sector}
% In order to obtain the mass matrix for the charged sector, we compute the second derivative of Higgs potential with respect to each charged field, they are: $\Phi_{C_i}:= \{ \chi^{\pm}\,,\,\rho^{\pm}\,,\,\eta^{\mp}_2\,,\,\eta^{\mp}_3 \} $.

\begin{eqnarray}
M^2_{C_{ij}}=\frac{\partial^2 V_H}{\partial\Phi_{C_i}\partial\Phi_{C_j}}\bigg|_{\Phi_{C_j}=0}
\label{charged-2ndderiv}
\end{eqnarray}

\noindent in the basis $\Phi_{C_i}=\{\chi^{\pm}\,,\,\eta^{\pm}_3\,,\,\rho^{\pm}\,,\,\eta^{\pm}_2\}$, which can be separated into two independent $2 \times 2$ blocks in the $\{\chi^{\pm}\,,\,\eta^{\pm}_3\}$ and $\{\rho^{\pm}\,,\,\eta^{\pm}_2\}$ basis, as shown in Eqs. (\ref{app:chargedmixing})-(\ref{app:charged-block}) in Appendix \ref{mat_masa_cargado}. The diagonalization of these matrices lead us to the following mass eigenstates:

\begin{eqnarray}
\begin{pmatrix}G_1^{\pm} \\ H_1^{\pm}\end{pmatrix} &=& \mathbb{R}_1\begin{pmatrix}\chi^{\pm} \\ \eta_3^{\pm}\end{pmatrix}, \ \ \mathbb{R}_1 = \begin{pmatrix}c_{\alpha} & e^{-i\delta}s_{\alpha}\\ -e^{i\delta}s_{\alpha} & c_{\alpha} \end{pmatrix}, \label{charged-mass-stat}\\
\begin{pmatrix}G_2^{\pm} \\ H_2^{\pm}\end{pmatrix} &=& \mathbb{R}_2\begin{pmatrix}\rho^{\pm} \\ \eta_2^{\pm}\end{pmatrix}, \ \ \mathbb{R}_2 = \begin{pmatrix}c_{\theta} & s_{\theta}\\ s_{\theta} & c_{\theta} \end{pmatrix}
,\end{eqnarray}
where the mixing angles and the complex phase are defined as: 

\begin{eqnarray}
s_{\alpha} = \sin{\alpha}=-\frac{|w|}{u}\;;&&\;\;\; |w|=\frac{w_1}{v_1}|v|\;,\\
\label{delta_rel} e^{i\delta}=\frac{v_1+i v_2}{|v|} \;;&&\;\;\;|v|=\sqrt{v_1^2+v_2^2}\;,\\
\label{theta_rel} s_{\theta}=\sin{\theta}=-\frac{|w|}{v_{ew}}\;;&&\;\;\; v_{ew}=\sqrt{|w|^2+|v|^2} \label{theta-mixing}
,\end{eqnarray}
with $|w|,\,|u|,\,|\nu|$ the norm of the VEV's and  $v_{ew}\approx246$ GeV the electroweak breaking scale. The fields $G_{1,2}^{\pm}$ are the corresponding massless Goldstone bosons associated to the massive $W^{\pm}$ and $V^{\pm}$ gauge bosons. The other charged Higgs bosons remains as physical particles with the following squared masses:

\begin{eqnarray}
\nonumber m_{H_1}^2&=&\frac{1}{2}\left(l_{7} + \sqrt{2} f \frac{v_{1}}{u w_1}\right)\left( u^{2} + \frac{w_1^2}{v_1^2}\left(v_{1}^{2} +  v_{2}^{2}\right) \right)\\
&\approx& \frac{1}{2}l_7 \, u^2,\\
\nonumber m_{H_2}^2&=&\frac{1}{2}\left(l_9(v_1^2+v_2^2)+\sqrt{2}f\frac{u v_1}{w_1} \right)\left(1+\frac{w_1^2}{v_1^2}\right)\\
&\approx&\frac{f}{\sqrt{2}}\left(\frac{u (v_1^2+w_1^2)}{v_1 w_1} \right)\label{h2charged-mass}
,\end{eqnarray}
where the approximations can be considered if we assume that the first transition occurs at much larger scale than the second one, i.e., if $u \gg v_1,v_2, w_1, w_2$.

For the neutral components, we have:

\begin{equation}
M^2_{N_{ij}}=\frac{\partial^2 V_H}{\partial\Phi_{N_i}\partial\Phi_{N_j}}\bigg|_{\Phi_{N_j}=0},\label{neutral2ndderiv}
\end{equation}
which lead us to two independent mass matrices in the basis $\{\xi_{\chi_2}, \zeta_{\chi_2},\xi_{\rho_3},\zeta_{\rho_3}\}$ and $\{\zeta_{\chi_3}, \zeta_{\rho_2}, \zeta_{\eta}, \xi_{\rho_2},\xi_{\eta},\xi_{\chi_3}\}$ written in Appendices \ref{mat_masa_neutro} and \ref{mat_masa_neutro_b}, where we can see that mixing terms between CP-even ($\xi_i $) and CP-odd ($\zeta _i$) fields emerge, which will induce CP violation. For the $4\times 4$ mass matrix, we make a block diagonalization by organizing the matrix in $2\times 2$ submatrices, as shown in equations (\ref{4x4neutral}) - (\ref{c-submatrix}), which exhibits a hierarchy structure that allow us apply a recursive expansion to obtain the set of eigenvalues and eigenstates, as shown in App.  \ref{mat_masa_neutro}. This procedure results in the following mass eigenstates:   

\begin{equation}
\begin{pmatrix}
G_1^{0} \\ G_2^0 \\ H_1^{0} \\ H_2^0 
\end{pmatrix}=\mathbb{R}_3
\begin{pmatrix}
\xi_{\chi^0_2} \\ \zeta_{\chi_2^0} \\ \xi_{\rho_3^0} \\ \zeta_{\rho_3^0} 
\end{pmatrix}, 
\label{neutral-1mass}
\end{equation}

\begin{equation}
\mathbb{R}_3 = \begin{pmatrix}c_{\Omega} & 0 & s_{\Omega}c_{\delta} & s_{\Omega}s_{\delta} \\ 0 & c_{\Omega} & s_{\Omega}s_{\delta} & -s_{\Omega}c_{\delta} \\ -s_{\Omega}c_{\delta} & -s_{\Omega}s_{\delta} & c_{\Omega} & 0 \\ -s_{\Omega}s_{\delta} & s_{\Omega}c_{\delta} & 0 & c_{\Omega} \end{pmatrix},
\end{equation}
where the mixing angle is:

\begin{equation}
s_{\Omega}\equiv\sin{\Omega}=- \frac{|v|}{u}.
\end{equation}

The neutral Higgs bosons have the following squared masses:

\begin{eqnarray}
m^2_{H_1^0} \approx \frac{u^2}{2} ( l_{8} + 2l_{10}) \label{h10-mass}\\
m^2_{H_2^0}\approx \frac{u^2}{2} ( l_{8} - 2l_{10}) \label{h20-mass}.
\end{eqnarray}

For the other neutral matrix, we also define a block structure, as shown in Eq. (\ref{6x6neutral}), where we identify three scales: very light ($\sim v_{ew}^3/u$), light ($\sim v_{ew}^2$) and heavy ($\sim v_{ew}u$ and $u^2$). This hierarchy structure allow us to obtain the diagonalization of the $6\times 6$ matrix, as shown in Appendix \ref{mat_masa_neutro_b}, obtaining the following mass eigenstates:

\begin{eqnarray}
G_3^0&\approx&\zeta_{\chi_3^0} \label{golds-3} \\
H_3^0&\approx&\xi_{\chi_3^0} \label{neutral-Higgs3}  \\
\begin{pmatrix}
G_4^{0} \\ H_4^0 \\ H_5^{0} \\ h^0 
\end{pmatrix}
&\approx&
\mathbb{R}_4\begin{pmatrix}\zeta_{\rho^0_2} \\ \zeta_{\eta^0} \\ \xi_{\rho_2^0} \\ \xi_{\eta^0} \end{pmatrix}, \label{eigenvect}
\end{eqnarray}
with

\begin{equation}
\mathbb{R}_4 = \begin{pmatrix}c_{\theta}c_{\delta} & -s_{\theta}c_{\delta} & -c_{\theta}s_{\delta} & -s_{\theta}s_{\delta} \\ -s_{\theta}c_{\delta} & -c_{\theta}c_{\delta} & s_{\theta}s_{\delta} & -c_{\theta}s_{\delta} \\ s_{\theta}s_{\delta} & c_{\theta}s_{\delta} & s_{\theta}c_{\delta} & -c_{\theta}c_{\delta} \\ c_{\theta}s_{\delta} & -s_{\theta}s_{\delta} & -c_{\theta}c_{\delta} & -s_{\theta}c_{\delta} \end{pmatrix}
,\end{equation}
and $\theta $ the same mixing angle obtained in Eq. (\ref{theta-mixing}). In the above spectrum, we find two Goldstone bosons ($G_{3,4}^0$) and four Higgs bosons with the following masses

\begin{eqnarray}
M^2_{H^0_3}&\approx &2 l_{1} u^{2} \label{heavy-higgsMass} \\ 
M^2_{H^0_4}&\approx &M^2_{H^0_5}=\frac{ f }{\sqrt{2}}\left(\frac{v_{1}^{2} + w_{1}^{2}}{v_{1} w_{1}}\right)u\\
M^2_{h^0} &\approx& \frac{1}{2}\left(\frac{v_1^2+v_2^2}{v_1^2}\right)\left(\frac{l_c w_1^4-2l_bw_1^2v_1^2-l_av_1^4}{v_1^2+w_1^2} \right) \label{light-higgsMass} 
\end{eqnarray}  
where: 

\[l_a=l_4^2-4l_1l_2\;\;,\;\;-l_c=l_5^2-4l_1l_3\;\;,\;\;l_b=l_4l_5-2l_1l_6.\]

We observe that one of the Higgs bosons ($h^0$) has mass at the electroweak scale, which we will identify with the observe one at LHC.

In summary, Table  \ref{mass_phisical_scalar} shows the physical scalar spectrum of the model.

\begin{table}[h]
\centering
\caption{Physical Scalar Spectrum }
\label{mass_phisical_scalar}
\begin{tabular}{c | c }
\hline
\hline
Scalar Boson & Squared Mass \\
\hline
$G_{1,2}^{\pm}$ & 0 \\
$G_{1,2,3,4}^0$ & 0 \\
$H_1^{\pm}$ & $\frac{1}{2}l_7 \, u^2$\\
$H_2^{\pm}\;,\;H_4^{0}\;,\;H_5^{0}$ & $\frac{f}{\sqrt{2}}u\left(\frac{\nu_{ew}^2}{|v||w|} \right)$ \\
$H_1^{0}$ & $\frac{1}{2}\left(l_8+l_{10}\right)\,u^2$\\
$H_2^{0}$ & $\frac{1}{2}\left(l_8-l_{10}\right)\,u^2$\\
$H_3^{0}$ & $2\, l_1 \, u^2$\\
$h^{0}$ & $\frac{1}{2}\left(\frac{l_c \,|w|^4\,-2\,l_b\,|w|^2|v|^2\,-l_a\,|v|^4}{\nu_{ew}^2} \right)$\\
\hline  
\hline
\end{tabular}
\end{table}

\subsection{Physical Spectrum - Vector Sector}

The interactions with the scalar fields are incorporated through the kinetic Higgs Lagrangian.

\begin{equation}\label{kinetic_eq2}
\resizebox{.42 \textwidth}{!}{
$
\mathcal{L}_K=(D_{\mu}\chi)^{\dagger}(D^{\mu}\chi)+(D_{\mu}\rho)^{\dagger}(D^{\mu}\rho)+(D_{\mu}\eta)^{\dagger}(D^{\mu}\eta)
$}
,\end{equation}
where $D_{\mu}$ is the covariant derivative, defined in the equation (\ref{derivative_331}). The mass matrix can be obtained through:

\begin{equation}
M^2_{V_{ij}}=\frac{\partial^2 \mathcal{L}_K}{\partial\ V_{i}\partial \Phi_{j}}\bigg|_{\Phi_{j}=0}
,\end{equation}
where $\Phi_i$ are the scalar fields in weak basis and $V_i$ are the vector fields. The charged bosons and the complex neutral bosons ($V^{0}$), are already in mass states with masses:

\begin{eqnarray}
M^2_{W^{\pm}}&=&\frac{g_L^2}{4}v_{ew}^2\\
M^2_{V^{\pm}}&=&\frac{g_L^2}{4}\left( u^2+|w|^2\right)\\
M^2_{V^{0},\overline{V}^{0}}&=&\frac{g_L^2}{4}\left( u^2+|v|^2\right) 
.\end{eqnarray}

Since $v_{ew}^2 = |v|^2+|w|^2 \ll u^2$, we obtain the hierarchy $M_{W^{\pm}} \ll M_{V}$.
On the other hand, the other neutral gauge bosons exhibit mixing mass terms. This matrix is singular, ensuring one massless boson (the photon). In 331 models, the electroweak gauge bosons rotate into mass eigenstates as:

\begin{eqnarray}
A_{\mu } &=&S_{W}W_{\mu }^{3}+C_{W}\left(\frac{1}{\sqrt{3}}T_{W}W_{\mu }^{8}+f[T_W]B_{\mu }\right) ,  \notag \\
Z_{\mu } &=&C_{W}W_{\mu }^{3}-S_{W}\left(\frac{1}{\sqrt{3}}T_{W}W_{\mu }^{8}+f[T_W]B_{\mu }\right) ,  \notag \\
Z_{\mu }^{\prime } &=&-f[T_W]W_{\mu }^{8}+\frac{1}{\sqrt{3}}T_{W}B_{\mu },
\end{eqnarray}%
where the Weinberg angle is defined as 

\begin{equation}
\sin \theta _{W}=S_{W}=\frac{g^{\prime }}{\sqrt{g^{2}+4g^{\prime 2}/3}},
\label{weinberg}
\end{equation}%
and $f[T_W]=\sqrt{1-T_W^2/3}$. However, in the above rotation, there is still a small mixing in the basis $(Z,Z')$, which diagonalize through a rotation with angle $\varphi$ given approximately by:

\begin{eqnarray}
\sin \varphi = s_{\varphi}\approx \frac{\sqrt{3}f[T_W]}{4C_W^3}\left(\frac{|v|^2C_{2W}-|w|^2}{u^2}\right),
\end{eqnarray}
so that the mass eigenstates rotate as:

\begin{equation}
\begin{pmatrix} A_{\mu} \\ Z^1_{\mu} \\ Z^2_{\mu} \end{pmatrix}=\mathbb{R}_{bos}\begin{pmatrix} W^{3}_{\mu} \\ W^{8}_{\mu} \\ B_{\mu} \end{pmatrix}\;,
\end{equation}
where the rotation matrix is defined as:

\begin{equation}
\resizebox{.47 \textwidth}{!}{
$
\mathbb{R}_{bos}=\begin{pmatrix} S_W & \frac{1}{\sqrt{3}}S_W & f[T_W]C_W \\ C_Wc_{\varphi}  & f[T_W]s_{\varphi}-\frac{1}{\sqrt{3}}S_WT_Wc_{\varphi}& -f[T_W]S_Wc_{\varphi}-\frac{1}{\sqrt{3}}T_Ws_{\varphi} \\ C_Ws_{\varphi} & -f[T_W]c_{\varphi}-\frac{1}{\sqrt{3}}S_WT_Ws_{\varphi} & \frac{1}{\sqrt{3}}T_Wc_{\varphi}-f[T_W]S_Ws_{\varphi} \end{pmatrix}.
$}
\end{equation}
%revisar porque el profe en la tesis me sugirio aqui la matriz transpuesta y revisar signos
%The $Z^1_{\mu}$ and $Z^{2}_{\mu}$ are not the same bosons as the usual $Z$ and $Z'$ in other extended models, because there are a resulting mixing between them. 
$Z^1_{\mu}$ can be identified as the phenomenological neutral weak boson and $Z^2_{\mu}$ is a new heavy physical boson. As we can see, the angle $\varphi$ that mix the components to obtain $Z^1_{\mu}$ and $Z^{2}_{\mu}$ is suppressed as $s_{\varphi}\propto v_{ew}/u^2$. In the limit without mixing, $Z^1_{\mu} \rightarrow Z$ and $Z^{2}_{\mu} \rightarrow Z'$.\\

As in the SM, we obtain that $C_{W}=\frac{M^2_{W}}{M^2_{Z}}$, which allow us to identify the angle $\theta _W$ with the Weinberg angle. Table  \ref{mass_phisical_boson} shows the physical vector bosons and their masses. 

\begin{table}[h]
\centering
\caption{Physical Vectorial Fields Spectrum }
\label{mass_phisical_boson}
\begin{tabular}{c | c }
\hline
\hline
Vector Boson & Squared Mass \\
\hline
$A_{\mu}$ & 0 \\
$W^{\pm}_{\mu}$ & $\frac{g_L^2}{4}\nu_{ew}^2$ \\
$V^{\pm}_{\mu}$ & $\frac{g_L^2}{4}\left( u^2+|w|^2\right)$ \\
$V^{0}_{\mu},\overline{V^0}_{\mu}$ & $\frac{g_L^2}{4}\left( u^2+|\nu|^2\right)$ \\
$Z_{\mu}$ & $\frac{g_L^2}{4C^2_{W}}v_{ew}^2$\\
$Z'_{\mu}$ & $\frac{g_X^2}{3\,T^2_{W}}u^2$\\
\hline  
\hline
\end{tabular}
\end{table}

%By transforming both, the scalar and vector boson basis into their physical mass eigenstates into the kinetic Lagrangian, we can obtain new CP violating coupling terms with phenomenological consequences.

\subsection{Physical Spectrum - Yukawa Sector}

Due to the $Z_2$ symmetry which was chosen in the Higgs potential for the model, some terms in the Yukawa Lagrangian must be removed, spoiling the mass acquisition of the fermions. In particular, introducing this symmetry, all the light fermions (those terms that couple with $\rho$ and $\eta$) become massless.\\

Thus, it is necessary to include additional $Z_2$ symmetries in the quarks sector in order to guarantee mass for all the fermionic fields. We choose the following symmetries:\\

\paragraph{Right-handed Sector}:
\begin{eqnarray}
u^{(1)}_R \rightarrow  u^{(1)}_R \;\;\;&,&\;\;\; d^{(1)}_R \rightarrow  -d^{(1)}_R \;,\\
u^{(2,3)}_R \rightarrow  -u^{(2,3)}_R \;\;\;&,&\;\;\; d^{(2,3)}_R \rightarrow  d^{(2,3)}_R \;,
\end{eqnarray}

\paragraph{Left-handed Sector}:
\begin{eqnarray}
Q^{(1)}_L &\rightarrow&  Q^{(1)}_L \;,\\
Q^{(2,3)}_L &\rightarrow&  -Q^{(2,3)}_L \;,
\end{eqnarray}

Taking into account the above symmetries, it is possible to set the Yukawa Lagrangian. After that and computing in the vacuum states of scalar fields, we obtain the following mass matrix for the up sector:

\begin{equation*}
\mathbb{M}^2_U=\begin{blockarray}{ccccc}
U^{(1)}_R & U^{(2)}_R & U^{(3)}_R & J^{(3)}_R \\
\begin{block}{(ccc|c)c}
  0 & \Gamma^{(_U,\eta)}_{1,2}\,w & \Gamma^{(_U,\eta)}_{1,3}\,w & 0 & \,\bar{U}^{(1)}_L \\
  \Gamma^{(_U,\eta)}_{2,1}\,w & 0 & 0 & \Gamma^{(_J,\eta)}_{2,2}\,w & \;\bar{U}^{(2)}_L \\
  \Gamma^{(_U,\rho)}_{3,1}\,\nu^{\ast} & 0 & 0 & \Gamma^{(_J,\rho)}_{3,3}\,\nu^{\ast} & \;\bar{U}^{(3)}_L \\ 
---&---&---&---& \\
  0 & \Gamma^{(_U,\chi)}_{3,2}\,u & \Gamma^{(_U,\chi)}_{3,3}\,u & 0 & \bar{J}^{(3)}_L \\
\end{block}
\end{blockarray}
\end{equation*}
and the matrix for the down sector:
\begin{equation*}
\mathbb{M}^2_D=\begin{blockarray}{cccccc}
D^{(1)}_R & D^{(2)}_R & D^{(3)}_R & J^{(1)}_R & J^{(1)}_R \\
\begin{block}{(ccc|cc)c}
 \Gamma^{(_D,\rho)}_{1,1}\,\nu & 0 & 0 & 0 & \Gamma^{(_J,\rho)}_{1,2}\,\nu & \bar{D}^{(1)}_L \\
 0 & \Gamma^{(_D,\rho)}_{2,2}\,\nu & \Gamma^{(_D,\rho)}_{2,3}\,\nu & \Gamma^{(_J,\rho)}_{2,1}\,\nu & 0 & \bar{D}^{(2)}_L \\
 0 & \Gamma^{(_D,\eta)}_{3,2}\,w^{\ast} & \Gamma^{(_D,\eta)}_{3,3}\,w^{\ast} & \Gamma^{(_J,\eta)}_{3,1}\,w^{\ast} & 0 & \bar{D}^{(3)}_L \\
---&---&---&---&---& \\
 0 & \Gamma^{(_D,\chi)}_{1,2}\,u & \Gamma^{(_D,\chi)}_{1,3}\,u & \Gamma^{(_J,\chi)}_{1,1}\,u & 0 & \bar{J}^{(1)}_L \\
 \Gamma^{(_D,\chi)}_{2,1}\,u & 0 & 0 & 0 & \Gamma^{(_J,\chi)}_{2,2}\,u & \bar{J}^{(2)}_L \\
\end{block}
\end{blockarray}
 \end{equation*}

In general, those coupling parameters ($\Gamma$) are complex. Then we can obtain the physical states with a bi-unitary transformation.\\

After diagonalizing, the fields are obtained in mass eigenstates, however, this sector exhibits many phenomenological aspects as the problem of Flavor Changing Neutral Currents (FCNC's) that must be controlled in this model, the hierarchy of quarks masses and the violation effects through the CKM matrix. These aspects are outside of the central purpose of this work but they can be considered for a later research.

\section{Conclusions}

We study the most general scalar potential for the 331 model with a parameter $\beta=1/\sqrt{3}$. Using the Gauge symmetries of the model we rotate the scalar fields in order to eliminate non-physical phases to remain exclusively with the physical phases\footnote{In the appendix \ref{mat_trans_su3}, it is illustrated in detail; how to use the Gauge symmetries of $SU(3)\otimes U(1)$ to rotate the complex phases and stay with only the physical phases.} which allow generating a CP violation mechanism in the scalar sector.\\

We find that taking the model with three Higgs triplets only two complex physical phases are required. We consider a particular phenomenological scenario in which these phases are fixed in the scalar doublets associated with the weak transition ($2^{\text{nd}}$.T.), where the CP violation would exhibit at low energies on an electroweak scale. Although it is possible to consider the scenario where one of these phases is on the electroweak scale while the other one is in the TeV breaking scale, this scenario will be considered in a later work.\\

For our case, we find the mass eigenstates spectrum of the even and odd fields, in both the neutral and the charged sector of the model. From the conditions of the minimum potential, it is possible to eliminate one of the complex phases and rewrite the mass eigenstates and rotations according to only one of the complex phases. This phase only appears in the rotation matrix of the charged Higgs sector. The neutral sector is free of this complex phase. The presence of CP violation is also reflected in the mixture of even scalar fields with the odd fields. However, from LHC data in the Higgs decay in two photons, there is a very small gap in which the Higgs boson has a mixture from the odd sector \cite{Brod:2013cka,Inoue:2014nva,Keus:2015hva}, therefore the mixture of the odd's with the even's it is very restricted by LHC data.\\

We also consider additional global symmetries where it is shown that complex phases can be reabsorbed into the parameters of the scalar potential and therefore in those particular scenarios there is no CP violation, leaving only the global symmetry scenario $Z_2$ which allows a framework with spontaneous CP violation.

%===============================//============================
%  ANEXOS  ///////////////////////////////////////////////////////////////////////////
%===============================//============================

\onecolumngrid
\appendix

%===============================//============================
\section{Matrix Transformation for SU(3) and rotations}\label{mat_trans_su3}
%===============================//============================

The spectral theorem for smooth functions of a Hermitian matrix with eigenvalues $m_j$ allow the expansion:

\begin{eqnarray}
f(\mathbb{M})=\sum_{j=1}^{N}f(m_j)\mathbb{P}_j,
\end{eqnarray}
with

\begin{eqnarray}
\mathbb{P}_j=\prod _{i\neq j} \frac{\mathbb{M}-m_i}{m_j-m_i}
\end{eqnarray}
and expanded as in Equation (\ref{productoria}), from where \cite{weigert1997baker,curtright2015elementary}: %and spectral theorem. One function can be written as a sum of powers of the matrix element:
\begin{equation}
 f(\mathbb{M})=\sum_{n=0}^{N-1}\left(\sum _{j=1}^N P_{jn}f(m_j)\right)\mathbb{M}^n=\sum_{n=0}^{N-1}\mathrm{f}_n(\mathbb{M})\mathbb{M}^n
,\end{equation}

for example, for our purpose, the exponential function ($f(\mathbb{M}):=e^{i\mathbb{M}}$), can be written as:

\begin{equation}
e^{i\mathbb{M}}=\sum_{n=0}^{N-1}\left(\sum _{j=1}^N P_{jn}e^{im_j}\right)\mathbb{M}^n
\label{exp_pot_2}
\end{equation}

On the other hand, the hermitian matrix $\mathbb{M}$ can be written as the product between the rotation angle ($\Omega$) with the group generators as $\mathbb{M}=\vec{\Omega} \cdot \vec{T}$. By writing ($\Omega = \alpha \hat{n}_{\alpha}$), we can rewrite the matrix as ($\mathbb{M}=\alpha(\hat{n}_{\alpha}\cdot\vec{T})=\alpha \hat{\mathbb{M}}$). For the $SU(3)$ group, we have that $\hat{n}_{\alpha}=\{n_1, ..., n_8\}$, so that the powers of the matrix $\hat{\mathbb{M}}$ are:

\begin{eqnarray}
\hat{\mathbb{M}}^0&=&\mathbb{I} \label{identity_m} \\
(\hat{n} \cdot \vec{T})=\hat{\mathbb{M}}^1&=&\frac{1}{2}
\begin{pmatrix} n_3+\frac{1}{\sqrt{3}}n_8 & n_1-in_2 & n_4-in_5 \\ n_1+in_2 & -n_3+\frac{1}{\sqrt{3}}n_8 & n_6-in_7 \\ n_4+in_5 & n_6+in_7 & -\frac{2}{\sqrt{3}}n_8 \end{pmatrix}=
\begin{pmatrix} a & b & c \\ b^* & d & e \\ c^* & e^* & -(a+d) \end{pmatrix} \label{linear_m} \\
\hat{\mathbb{M}}^2&=&
\begin{pmatrix} a^2+b^2+c^2 & b(a+d)+ce^* & -cd+be \\ b^*(a+d)+c^*e & d^2+b^2+e^2 & -ed+cb^* \\ -c^*d+b^*e^* & -e^*d+c^*b & c^2+e^2+(a+d)^2 \end{pmatrix} \label{cuadratic_m}
,\end{eqnarray}

\noindent changing the space of parameters from 8 real parameters ($ n_{1,2, ..., 8}$) to 3 complex parameters ($ b, c, e $) and 2 real parameters ($ a, d $). From the spectral expansion in Equation (\ref{exp_pot_2}) for $SU(3)$, the exponential can be written as:

\begin{equation}
e^{i\mathbb{M}}=\mathrm{f}_0\mathbb{I}+\mathrm{f}_1\hat{\mathbb{M}}+\mathrm{f}_2\hat{\mathbb{M}}^2
\end{equation}
where \cite{curtright2015elementary}
\begin{eqnarray}
\mathrm{f}_0&=&\sum_{k=0,1,2}-\frac{1}{3}(1+2\cos(2\phi + 4\pi k/3))\mathcal{F}\\
\mathrm{f}_1&=&\sum_{k=0,1,2}\frac{2}{\sqrt{3}}\sin(\phi + 2\pi k/3)\mathcal{F}\\
\mathrm{f}_2&=&\sum_{k=0,1,2}\mathcal{F}
\end{eqnarray}
and
\[ \mathcal{F}=\frac{e^{\frac{2}{\sqrt{3}}i\alpha\sin(\phi + 2\pi k/3)}}{1-2\cos(2\phi + 4\pi k/3)}\;\;,\;\;\phi= - \frac{1}{3}\arcsin \left(\frac{3\sqrt{3}}{2}\text{det}(\mathbb{M}) \right).\]

\noindent Using equations (\ref{identity_m}, \ref{linear_m} and \ref{cuadratic_m}), we obtain the transformation in matrix form ( $e^{i\mathbb{M}}=\mathbb{T}$) as:

\begin{equation}
\mathbb{T}=
\begin{pmatrix} \mathbb{T}_{11} & \mathrm{f}_1b+\mathrm{f}_2(b(a+d)+ce^*) & \mathrm{f}_1c+\mathrm{f}_2(-cd+be) \\ 
\mathrm{f}_1b^*+\mathrm{f}_2(b^*(a+d)+c^*e) & \mathbb{T}_{22}& \mathrm{f}_1e+\mathrm{f}_2(-ed+cb^*) \\ 
\mathrm{f}_1c^*+\mathrm{f}_2(-c^*d+b^*e^*) & \mathrm{f}_1e^*+\mathrm{f}_2(-e^*d+c^*b) & \mathbb{T}_{33} \end{pmatrix}
\label{app:matrix-repres}
\end{equation}
where
\begin{eqnarray}
\mathbb{T}_{11}&=&\mathrm{f}_0+\mathrm{f}_1a + \mathrm{f}_2(a^2+b^2+c^2)\nonumber \\
\mathbb{T}_{22}&=&\mathrm{f}_0+\mathrm{f}_1d+\mathrm{f}_2(d^2+b^2+e^2) \nonumber \\
\mathbb{T}_{33}&=&\mathrm{f}_0-\mathrm{f}_1(a+d)+\mathrm{f}_2(c^2+e^2+(a+d)^2).
\end{eqnarray}

The matrix (\ref{app:matrix-repres}) can be written as in equation (\ref{mat_rot_repam}),
\begin{equation}
\mathbb{T}=
\begin{pmatrix} T_1 & m_1+n_1 & m_2 + n_2 \\ m_1 - n_1 & T_2 & m_3 + n_3 \\ m_2 - n_2 & m_3 - n_3 & T_3 \end{pmatrix}
\end{equation}
if we define:

\begin{eqnarray}
m_1&=&[\mathrm{f}_1+\mathrm{f}_2(a+d)]\text{Re}(b)+\mathrm{f}_2\text{Re}(ce^*) \nonumber \\
n_1&=&[\mathrm{f}_1+\mathrm{f}_2(a+d)]i\text{Im}(b)+\mathrm{f}_2i\text{Im}(ce^*)\nonumber \\
m_2&=&[\mathrm{f}_1-\mathrm{f}_2d]\text{Re}(c)+\mathrm{f}_2\text{Re}(be^*)\nonumber \\
n_2&=&[\mathrm{f}_1-\mathrm{f}_2d]i\text{Im}(c)+\mathrm{f}_2i\text{Im}(be^*)\nonumber \\
m_3&=&[\mathrm{f}_1-\mathrm{f}_2d]\text{Re}(e)+\mathrm{f}_2\text{Re}(cb^*)\nonumber \\
n_3&=&[\mathrm{f}_1-\mathrm{f}_2d]i\text{Im}(e)+\mathrm{f}_2i\text{Im}(cb^*)\nonumber \\
T_1&=&\mathbb{T}_{11},\;\;\;T_2=\mathbb{T}_{22},\;\;\;T_3=\mathbb{T}_{33}
\label{app:reparam}
\end{eqnarray}

We can use this matrix representation $\mathbb{T}$, to make a $SU(3)$ rotation that eliminates one VEV from ($\rho$). On the other hand, if we apply an element of $SU(3)$ over the VEV of the scalar field $\eta$, the unbroken element must remain invariant, i.e., 

\begin{equation}
\langle\eta'\rangle=\mathbb{T}\langle\eta\rangle
.\end{equation}

\noindent Using explicit matrix representation we have:
\begin{equation}
\begin{pmatrix} \omega' \\ 0 \\ 0 \end{pmatrix} = 
\begin{pmatrix} T_1 & m_1+n_1 & m_2 + n_2 \\ m_1 - n_1 & T_2 & m_3 + n_3 \\ m_2 - n_2 & m_3 - n_3 & T_3 \end{pmatrix}
\begin{pmatrix} v_{\eta} \\ 0 \\ 0 \end{pmatrix}
,\end{equation}
and considering the mentioned invariance, we obtain the following relations;
\begin{equation}
T_1=\frac{\omega'}{v_{\eta}}, \;\;\;\; m_1=n_1, \;\;\;\; m_2=n_2
,\end{equation}
obtaining the following form:

\begin{equation}\label{536}
\mathbb{T}=
\begin{pmatrix} \frac{\omega'}{v_{\eta}} & 2m_1 & 2m_2  \\ 0 & T_2 & m_3 + n_3 \\ 0 & m_3 - n_3 & T_3 \end{pmatrix}
.\end{equation}

The scalar field ($\rho$) has two VEVs in general different from zero. However, we can rotate the triplet to obtain only one VEV in the second component. So, we demand that:

\begin{eqnarray}
\begin{pmatrix} 0 \\ v' \\ 0 \end{pmatrix} &=& 
\begin{pmatrix} T_1 &2 m_1 & 2m_2  \\ 0 & T_2 & m_3 + n_3 \\ 0 & m_3 - n_3 & T_3 \end{pmatrix}
\begin{pmatrix} 0 \\ v_{\rho_1} \\ v_{\rho_2}\end{pmatrix}
\end{eqnarray}
from where we obtain the following relations:

\begin{equation}
m_2=-m_1\frac{v_{\rho_1}}{v_{\rho_2}}, \;\;\;\; T_2=\frac{v'}{v_{\rho_1}}-(m_3+n_3)\frac{v_{\rho_2}}{v_{\rho_1}}, \;\;\;\; T_3=(n_3-m_3)\frac{v_{\rho_1}}{v_{\rho_2}}
.\end{equation}
which lead us to:

\begin{equation}
\mathbb{T}=
\begin{pmatrix} \frac{\omega'}{v_{\eta}} & 2m_1 & -2m_1\frac{v_{\rho_1}}{v_{\rho_2}} \\ 0 & \frac{v'}{v_{\rho_1}}-(m_3+n_3)\frac{v_{\rho_2}}{v_{\rho_1}}, & m_3 + n_3 \\ 0 & m_3 - n_3 & (n_3-m_3)\frac{v_{\rho_1}}{v_{\rho_2}}
\end{pmatrix}
.\end{equation}

Finally, the VEV of the scalar field $\chi$ transforms as:

\begin{eqnarray}
\begin{pmatrix} 0 \\ 0 \\ u' \end{pmatrix} &=& \mathbb{T}
\begin{pmatrix} 0 \\ 0 \\ v_{\chi}\end{pmatrix} \;\; 
,\end{eqnarray}
obtaining:

\begin{eqnarray}
m_1=0, \ \ \ m_3=-n_3, \ \ \ n_3 = \frac{v_{\rho_2}}{2v_{\rho_1}v_{\chi}}u',
\end{eqnarray}
so that the transformation matrix become:
% ($e^{-i2\Omega_{\rho}}$), then we have a new transformation matrix that satisfy: ($\mathbb{T}'''=\mathbb{T}''^{-2}$). Thus the $\mathbb{T}''$ determinant must be different from zero. It is:
%\begin{equation}
%\det\mathbb{T}''=\frac{2T_1m_2\nu'}{m_1\nu_1}(m_3-n_3) \;\; \rightarrow \;\; m_3\neq n_3
%.\end{equation}

%\noindent In a similar way than occurs with ($\eta$), this new matrix must satisfy;
%\begin{eqnarray}
%\nonumber \langle\chi'\rangle&=&\mathbb{T}'''\langle\chi\rangle\\
%\begin{pmatrix} 0 \\ 0 \\ u' \end{pmatrix} &=& \mathbb{T}'''
%\begin{pmatrix} 0 \\ 0 \\ u\end{pmatrix} \;\; 
%,\end{eqnarray}
%above transformation generates the following conditions;
%\begin{eqnarray}
% 0&=&\mathbb{T}'''_{23}u\\
% 0&=&-(m_3+n_3)\frac{m_1^2\nu_1^2}{2m_2^2\nu'^2}\left[ (m_3+n_3)\frac{m_1}{m_2}+\frac{\nu'}{\nu_1}+(m_3-n_3)\frac{m_2}{2m_1}\right]/(m_3-n_3)^2\\
%0&=&(m_3+n_3)\left[ (m_3+n_3)\frac{m_1}{2m_2}+\frac{2\nu'}{\nu_1}+(m_3-n_3)\frac{m_2}{2m_1}\right]
%.\end{eqnarray}

%\noindent We have the freedom to choose $m_3=-n_3$ as a solution in which the transformed fields give us the following results;
%\begin{equation}
%u'=\left(\frac{m_1}{m_2m_3}\right)^2u, \;\;\;\;\; T_1=-\frac{m_2m_3}{m_1} \;\;\;\;\;  \nu'=\left(\frac{\nu_2}{m_3}\right)^2\frac{1}{\nu_1}
.%\end{equation}

%Finally using above relations and (eq. \ref{536}), the final matrix transformation for $SU(3)_L$ gauge symmetry, that rotates the scalar field ($ \rho $) in order to leave only VEV in the second component and remain invariant the position for the other VEV's, is:
\begin{equation}
\mathbb{T}=
\begin{pmatrix} 
\frac{\omega'}{v_{\eta}}& 0 & 0 \\ 
0 & \frac{v'}{v_{\rho_1}} & 0 
\\ 0 & -\frac{v_{\rho_2}}{v_{\rho_1}v_{\chi}}u' & \frac{u'}{v_{\chi}}
\end{pmatrix}. 
\end{equation}

%===============================//============================
\section{Mass Matrices}
%===============================//============================

\subsection{Mass Matrix for the Charged Sector}\label{mat_masa_cargado}

Applying the second derivative as defined as in Eq. (\ref{charged-2ndderiv}), we obtain the following mass matrix for the charged sector:

\begin{equation}
M^2_C=\begin{pmatrix}M^2_{C1} & 0 \\ 0 & M^2_{C2} \end{pmatrix}
\label{app:chargedmixing}
\end{equation}
where;

\begin{eqnarray}
M^2_{C1}&=&
\begin{pmatrix}
\frac{1}{2} \, l_{7} w_{1}^{2} + \frac{l_{7} v_{2}^{2} w_{1}^{2}}{2 \, v_{1}^{2}} + \frac{\sqrt{2} f v_{1} w_{1}}{2 \, u} + \frac{\sqrt{2} f v_{2}^{2} w_{1}}{2 \, u v_{1}} & \frac{1}{2}\left( \frac{ l_{7} u w_{1}}{ v_{1}} + \sqrt{2} f \right) (v_{1} - i \, v_{2}) \\
 \frac{1}{2}\left( \frac{ l_{7} u w_{1}}{ v_{1}} + \sqrt{2} f \right) (v_{1} + i \, v_{2}) & \frac{1}{2} \, l_{7} u^{2} + \frac{\sqrt{2} f u v_{1}}{2 \, w_{1}}
\end{pmatrix} \\
M^2_{C2}&=&
\begin{pmatrix}
\frac{1}{2} \, l_{9} w_{1}^{2} + \frac{l_{9} v_{2}^{2} w_{1}^{2}}{2 \, v_{1}^{2}} + \frac{\sqrt{2} f u w_{1}}{2 \, v_{1}} & \frac{1}{2} \, l_{9} v_{1} w_{1} + \frac{l_{9} v_{2}^{2} w_{1}}{2 \, v_{1}} + \frac{1}{2} \, \sqrt{2} f u \\
\frac{1}{2} \, l_{9} v_{1} w_{1} + \frac{l_{9} v_{2}^{2} w_{1}}{2 \, v_{1}} + \frac{1}{2} \, \sqrt{2} f u & \frac{1}{2} \, l_{9} v_{1}^{2} + \frac{1}{2} \, l_{9} v_{2}^{2} + \frac{\sqrt{2} f u v_{1}}{2 \, w_{1}}
\end{pmatrix}\label{app:charged-block}.
\end{eqnarray}

\noindent Both matrices are singular, so we obtain at least two Goldstone bosons. By direct diagonalization of a $2\times2$ matrix, we find the eigenvectors and eigenvalues written in Eqs. (\ref{charged-mass-stat})-(\ref{h2charged-mass}). \\

\subsection{Mass Matrix for Neutral Sector - $\mathbb{M}_{a}$}\label{mat_masa_neutro}

Applying the second derivative from Eq. (\ref{neutral2ndderiv}) restricted to the basis $\Phi _{N_i}=(\xi_{\chi_2}\,,\,\zeta_{\chi_2}\,,\,\xi_{\rho_3}\,,\,\zeta_{\rho_3}$), we find the following blocks with hierarchy structure:

\begin{eqnarray}
\mathbb{M}_a=\begin{pmatrix}
A & B \\
B^T & C
\end{pmatrix},
\label{4x4neutral}
\end{eqnarray}
where

\begin{eqnarray}
A&=&\begin{pmatrix}
l_{10} (v_{1}^{2} - v_{2}^{2}) + |v|^2 \left(\frac{1}{2} \, l_{8} + \frac{\sqrt{2} f w_{1}}{2 \, u\, v_{1}} \right) & 2 l_{10} v_{1} v_{2} \\
2 l_{10} v_{1} v_{2} & -l_{10} (v_{1}^{2} - v_{2}^{2}) + |v|^2 \left(\frac{1}{2} l_{8} + \frac{\sqrt{2} f w_{1}}{2  u v_{1}} \right)
\end{pmatrix}  \\ \nonumber \\ \nonumber \\
B&=&\begin{pmatrix}
l_{10} u v_{1} + \frac{1}{2} l_{8} u v_{1} + \frac{1}{2} \sqrt{2} f w_{1} & -l_{10} u v_{2} + \frac{1}{2} l_{8} u v_{2} + \frac{\sqrt{2} f v_{2} w_{1}}{2 \, v_{1}} \\
l_{10} u v_{2} + \frac{1}{2}  l_{8} u v_{2} + \frac{\sqrt{2} f v_{2} w_{1}}{2  v_{1}} & l_{10} u v_{1} - \frac{1}{2}  l_{8} u v_{1} - \frac{1}{2}  \sqrt{2} f w_{1}
\end{pmatrix} \\ \nonumber \\ \nonumber \\
C&=&\begin{pmatrix}
l_{10} u^{2} + \frac{1}{2} l_{8} u^{2} + \frac{\sqrt{2} f u w_{1}}{2  v_{1}} & 0 \\
0 & -l_{10} u^{2} + \frac{1}{2} l_{8} u^{2} + \frac{\sqrt{2} f u w_{1}}{2 \, v_{1}}
\end{pmatrix}.\label{c-submatrix}
\end{eqnarray}

\noindent As we can see, matrix $C$ is proportional to the energy scale of the first transition ($C\sim u^2$), the matrix $B$ has intermediate energy scale ($B\sim u$) and finally, the matrix $A$ is at the electroweak scale ($A\sim \nu_{ew}$), so that:

\[C >> B >> A.\]
Therefore, the matrix (\ref{c-submatrix}) can be block diagonalized by a unitary rotation of the form \cite{Grimus_2000}:

\begin{eqnarray}
V=
\begin{pmatrix} 
1 & F \\ -F^T & 1 
\end{pmatrix},\label{unitary-neutral1}
\end{eqnarray}
where $1$ is the identity submatrix and $F$ is a submatrix that satisfy ($F<<1$). Keeping only up to linear terms on $F$, the matrix rotation has the form:

\[V^T\mathbb{M}_a V=\begin{pmatrix} A-BF^T-FB^T & B + AF - FC \\ B^T + F^TA - CF^T & C + B^TF+F^TB  \end{pmatrix}=\begin{pmatrix} \mathbb{D}_{a1} & 0 \\ 0 & \mathbb{D}_{a2}  \end{pmatrix}.\]
From the cancellation of the nondiagonal block, we obtain the solution

\[F\approx BC^{-1}.\]
Therefore, from the diagonal blocks, we obtain at dominant order that:

\begin{eqnarray}
\label{diag_block_1} \mathbb{D}_{a1}&\approx&A-B\,C^{-1}\,B^T\\
\label{diag_block_2} \mathbb{D}_{a2}&\approx&C
.\end{eqnarray}
By replacing each submatrix into the above results, we find that $\mathbb{D}_{a1}=0$, obtaining two Goldstone bosons, while $\mathbb{D}_{a2}$ is already diagonal, obtaining the two massive Higgs bosons written in (\ref{neutral-1mass}) and masses in (\ref{h10-mass}) and (\ref{h20-mass}).
 
\subsection{Neutral sector mass Matrix - $\mathbb{M}_{b}$}\label{mat_masa_neutro_b}

The mass matrix in the basis $\Phi_{N_i}=(\zeta_{\chi_3}\,,\,\zeta_{\rho_2}\,,\,\zeta_{\eta}\,,\,\xi_{\rho_2}\,,\,\xi_{\eta}\,,\,\xi_{\chi_3}$) appears as an independent matrix separated from the above matrix, $\mathbb{M}_{a}$, obtaining the following structure:

\begin{equation}
\mathbb{M}_{b}=
\begin{pmatrix} 
M_{b_{11}} & D & 0 \\ D^T & \mathbb{H} & E \\ 0 & E^T & M_{b_{66}}
\end{pmatrix},
\label{6x6neutral}
\end{equation}
where:

\begin{eqnarray}
M_{b_{11}}&=&\frac{\sqrt{2} f w_{1}}{2 \, u v_{1}} |v|^2 \label{11-mass}\\
M_{b_{66}}&=& 2 l_{1} u^{2} + \frac{\sqrt{2} f w_{1}}{2\,u v_{1} } |v|^2 \\
D&=& \left(  \frac{1}{2} \, \sqrt{2} f w_{1}, \frac{1}{2} \, \sqrt{2} f v_{1}, -\frac{\sqrt{2} f v_{2} w_{1}}{2 \, v_{1}}, \frac{1}{2} \sqrt{2} f v_{2} \right) \\
E^T&=& \left( \; l_{4} u v_{2} - \frac{\sqrt{2} f v_{2} w_{1}}{2 \, v_{1}} \;,\; -\frac{l_{5} u v_{2} w_{1}}{v_{1}} + \frac{\sqrt{2} f v_{2}}{2} \;,\; l_{4} u v_{1} - \frac{\sqrt{2} f w_{1}}{2} \;,\; l_{5} u w_{1} - \frac{\sqrt{2} f v_{1}}{2} \; \right) \\
\mathbb{H}&=&6
\begin{pmatrix}
 2 l_{2} v_{2}^{2} + \frac{\sqrt{2} f u w_{1}}{2 \, v_{1}} & -\frac{l_{6} v_{2}^{2} w_{1}}{v_{1}} + \frac{1}{2} \sqrt{2} f u & 2 \, l_{2} v_{1} v_{2} & l_{6} v_{2} w_{1} \\
 -\frac{l_{6} v_{2}^{2} w_{1}}{v_{1}} + \frac{1}{2}  \sqrt{2} f u & \frac{2  l_{3} v_{2}^{2} w_{1}^{2}}{v_{1}^{2}} + \frac{\sqrt{2} f u v_{1}}{2  w_{1}} & -l_{6} v_{2} w_{1} & -\frac{2 \, l_{3} v_{2} w_{1}^{2}}{v_{1}} \\
 2  l_{2} v_{1} v_{2} & -l_{6} v_{2} w_{1} & 2  l_{2} v_{1}^{2} + \frac{\sqrt{2} f u w_{1}}{2 \, v_{1}} & l_{6} v_{1} w_{1} - \frac{1}{2}  \sqrt{2} f u \\
 l_{6} v_{2} w_{1} & -\frac{2 \, l_{3} v_{2} w_{1}^{2}}{v_{1}} & l_{6} v_{1} w_{1} - \frac{1}{2} \sqrt{2} f u & 2  l_{3} w_{1}^{2} + \frac{\sqrt{2} f u v_{1}}{2  w_{1}}
\end{pmatrix}
\end{eqnarray}

By assuming that the VEVs of the triplets $\rho$ and $\eta$, and the cubic coupling constant $f$ are of the order of the electroweak scale (i.e. $f\sim v_{1,2}\sim w_{1,2}\sim v_{ew}$) while $\chi $ breaks the symmetry at a larger scale ($u\gg v_{ew}$), then the above blocks contains four energy scales:  $M_{b_{11}}$ is the lightest scale at $v_{ew}^3/u$, $D$ is the intermediate scale at $v_{ew}^2$, $E$ and $\mathbb{H}$ contains heavy terms at $v_{ew}u$, and $M_{b_{66}}$ is the heaviest scale at $u^2$. We may further reduce the matrix by putting the $v_{ew}u$ and $u^2$ scales into a single block, so that the mass matrix in (\ref{6x6neutral}) become:

\begin{equation}
\mathbb{M}_{b}=
\begin{pmatrix} 
M_{b_{11}} & B  \\ B^T & \mathbb{C}
\end{pmatrix},
\label{6x6neutral-2}
\end{equation}
where:

\begin{eqnarray}
B^T&=&
\begin{pmatrix}
D, & 0
\end{pmatrix} \\
\mathbb{C}&=&
\begin{pmatrix}
\mathbb{H} & E \\
E^T & M_{b66}
\end{pmatrix} \label{c-matrix2}
\end{eqnarray}
and $M_{b_{11}} \ll B \ll \mathbb{C}$. On the other hand, the exact $6\times6 $ matrix is singular with multiplicity two, containing two massless Goldstone bosons. One of this Goldstone bosons come from the submatrix $\mathbb{C}$, so that the other massless Goldstone must arise from the other blocks. Thus, we propose a block diagonalization of the form:

\begin{eqnarray}
U^T\mathbb{M}_{b}U=
\begin{pmatrix}
0 & 0 \\
0 & M
\end{pmatrix},\label{sublockdiag}
\end{eqnarray} 
where $U$ is an unitary rotation, which, as in (\ref{unitary-neutral1}), can be approximately written through a small transformation $F_{n\times m}$ of dimension $n\times m$  as:

\begin{eqnarray}
U=\begin{pmatrix}
1_{1\times 1} & F_{1\times 5} \\
-F^T_{5\times 1} & 1_{5\times 5}
\end{pmatrix},
\end{eqnarray}
so that at dominant order, we obtain: 

\begin{eqnarray}
M_{b_{11}}&\approx &F_{1\times 5}B^T \label{golds-1} \\
M &\approx &\mathbb{C} \label{golds-2}.
\end{eqnarray}
Eq. (\ref{golds-1}) represents a constraint over the components of the transformation $F_{1\times 5}$, while (\ref{golds-2}) tell us that, at dominant order, the sub-matrix $\mathbb{C}$ in (\ref{c-matrix2}) decouple from the other components, and can be block diagonalized separately. Taking into account that $\mathbb{H} \sim E \ll M_{b66}$, we can use the same recursive method as before, where a rotation matrix $R$ block diagonalize the matrix as:

\begin{eqnarray}
R^T\mathbb{C}R&=&
\begin{pmatrix}
d_1 & 0 \\
0 & d_2
\end{pmatrix},
\end{eqnarray}
with

\begin{eqnarray}
R=
\begin{pmatrix}
1_{4\times 4} & F_{4\times 1} \\
F^T_{1\times 4} & 1 _{1 \times 1}
\end{pmatrix}.
\end{eqnarray}
Thus, we obtain:

\begin{eqnarray}
F_{4\times 1}&\approx &E M_{b66}^{-1}, \nonumber \\
d_1&\approx& \mathbb{H}-E M_{b66}^{-1}E^T \nonumber \\
d_2&\approx& M_{b66}. 
\end{eqnarray}

The component $d_2$ gives directly the neutral Higgs boson written in Eq. (\ref{neutral-Higgs3}) with mass given by Eq. (\ref{heavy-higgsMass}), while $d_1$ is a new singular $4\times 4 $ matrix that can be diagonalized by analytic methods, obtaining the rotation matrix, mass eigenvectors and mass eigenvalues written in Eqs. (\ref{eigenvect}) - (\ref{light-higgsMass}). Finally, the Goldstone boson that arise in the $11$ component in (\ref{sublockdiag}) rotates through the transformation $F_{1\times5}$ that must accomplish the constraint from (\ref{golds-1}). As a first approximation, if we neglect this transformation, we obtain the Goldstone boson written in (\ref{golds-3})

%\pageref{#1}

\bibliography{bib_tesis_mod}

\begin{thebibliography}{10}

\bibitem{pisano1992}
{Pisano, F and Pleitez, V}.
\newblock { $SU(3) \otimes U(1)$ model for electroweak interactions}.
\newblock {\em {Physical Review D}}, {46}({1}):{410}, {1992}.

\bibitem{Frampton_1992}
{Frampton, P. H.}
\newblock {Chiral dilepton model and the flavor question}.
\newblock {\em {Phys. Rev. Lett.}}, {69}:{2889--2891}, {1992}.

\bibitem{buras2013B}
{Buras, Andrzej J and De Fazio, Fulvia and Girrbach, Jennifer and Carlucci,
  Maria V}.
\newblock {The anatomy of quark flavour observables in 331 models in the
  flavour precision era}.
\newblock {\em {Journal of High Energy Physics}}, {2013}({2}):{1--69}, {2013}.

\bibitem{ochoa2005}
{Ochoa, Fredy and Martinez, R}.
\newblock {Family dependence in
  $\mathrm{SU}(3)_C\otimes\mathrm{SU}(3)_L\times\mathrm{U}(1)_X$ models}.
\newblock {\em {Physical Review D}}, {72}({3}):{035010}, {2005}.

\bibitem{martinez2008}
{Martinez, R and Ochoa, F}.
\newblock {Mass-matrix ansatz and constraints on $B_s^0 \rightarrow
  \bar{B_s^0}$ mixing in 331 models}.
\newblock {\em {Physical Review D}}, {77}({6}):{065012}, {2008}.

\bibitem{montano2013B}
{Monta{\~n}o, J and P{\'e}rez, MA and Ram{\'\i}rez-Zavaleta, F and Toscano,
  JJ}.
\newblock {The $Z,Z' \rightarrow \gamma\gamma\gamma$ decays in the minimal 331
  model}.
\newblock {\em {Journal of Physics: Conference Series}}, {468}({1}):{012007},
  {2013}.

\bibitem{cogollo2012}
{Cogollo, D and Queiroz, FS and Teles, PR and de Andrade, A Vital}.
\newblock {Novel sources of Flavor Changed Neutral Currents in the $331_{RHN}$
  model}.
\newblock {\em {The European Physical Journal C}}, {72}({5}):{1--10}, {2012}.

\bibitem{promberger2007flavor}
{Promberger, Christoph and Schatt, Sebastian and Schwab, Felix}.
\newblock {Flavor-changing neutral current effects and CP violation in the
  minimal 331 model}.
\newblock {\em {Physical Review D}}, {75}({11}):{115007}, {2007}.

\bibitem{WEINBERG_1979}
{Weinberg, Steven}.
\newblock {Cosmological Production Of Baryons}.
\newblock {\em Phys. Rev. Lett.}, 42(13):850--853, 1979.

\bibitem{YOSHIMURA:1979aa}
{Yoshimura, Motohiko}.
\newblock {Origin Of Cosmological Baryon Asymmetry}.
\newblock {\em Phys. Lett. B}, 88(3-4):294--298, 1979.

\bibitem{Barr:1979ye}
Stephen~M. Barr, Gino Segre, and H.~Arthur Weldon.
\newblock {The Magnitude of the Cosmological Baryon Asymmetry}.
\newblock {\em Phys. Rev.}, D20:2494, 1979.

\bibitem{NANOPOULOS:1979aa}
D.V. Nanopoulos and {Weinberg, Steven}.
\newblock {Mechanisms For Cosmological Baryon Production}.
\newblock {\em Phys. Rev. D}, 20(10):2484--2493, 1979.

\bibitem{FRY_1980}
J.~N. Fry, K.~A. Olive, and M.~S. Turner.
\newblock {Evolution Of Cosmological Baryon Asymmetries}.
\newblock {\em Phys. Rev. D}, 22(12):2953--2988, 1980.

\bibitem{KOLB:1980aa}
E.~W. Kolb and S.~Wolfram.
\newblock {Baryon Number Generation In The Early Universe}.
\newblock {\em Nucl. Phys. B}, 172(1):224--284, 1980.

\bibitem{fukugita_1986}
{Fukugita, M and Yanagida, Tsutomu}.
\newblock {Baryogenesis without grand unification}.
\newblock {\em {Physics Letters B}}, {174}({1}):{45--47}, {1986}.

\bibitem{Luty:1992un}
M.~A. Luty.
\newblock {Baryogenesis via leptogenesis}.
\newblock {\em Phys. Rev.}, D45:455--465, 1992.

\bibitem{sakharov1967violation}
{Sakharov, Andrej Dmitrievich}.
\newblock {Violation of CP invariance, C asymmetry, and baryon asymmetry of the
  universe}.
\newblock {\em {JETP lett.}}, {5}:{24--27}, {1967}.

\bibitem{Manton:1983nd}
N.~S. Manton.
\newblock {Topology in the Weinberg-Salam Theory}.
\newblock {\em Phys. Rev.}, D28:2019, 1983.

\bibitem{Kuzmin:1985mm}
V.~A. Kuzmin, V.~A. Rubakov, and M.~E. Shaposhnikov.
\newblock {On the Anomalous Electroweak Baryon Number Nonconservation in the
  Early Universe}.
\newblock {\em Phys. Lett.}, 155B:36, 1985.

\bibitem{barbieri_2003}
{Barbieri, Riccardo and Creminelli, Paolo and Strumia, Alessandro and Tetradis,
  Nikolaos}.
\newblock {Baryogenesis through leptogenesis}.
\newblock {\em {Nuclear Physics B}}, {575}({1-2}):{61--77}, {2000}.

\bibitem{Giudice:2004aa}
G.~F. Giudice, A.~Notari, M.~Raidal, A.~Riotto, and A.~Strumia.
\newblock {Towards a complete theory of thermal leptogenesis in the SM and
  MSSM}.
\newblock {\em Nucl. Phys. B}, 685:89--149, May 2004.

\bibitem{nardi_2008}
{Davidson, Sacha and Nardi, Enrico and Nir, Yosef}.
\newblock {Leptogenesis}.
\newblock {\em {Physics Reports}}, {466}({4-5}):{105--177}, {2008}.

\bibitem{Covi:1996aa}
L.~Covi, E.~Roulet, and F.~Vissani.
\newblock {CP violating decays in leptogenesis scenarios}.
\newblock {\em Phys. Lett. B}, 384(1-4):169--174, September 1996.

\bibitem{kajantie_phase_1995}
{Kajantie, Keijo and Laine, Mikko and Rummukainen, K and Shaposhnikov, M}.
\newblock {The electroweak phase transition: a non-perturbative analysis}.
\newblock {\em {Nuclear Physics B}}, {466}({1-2}):{189--258}, {1996}.

\bibitem{SMCPV_93}
{Gavela, MB and Lozano, M and Orloff, J and Pene, O}.
\newblock {Standard Model CP-violation and Baryon asymmetry Part I: Zero
  Temperature}.
\newblock {\em {arXiv preprint hep-ph/9406288}}, {1994}.

\bibitem{SMCPV_quimbay_94}
{Gavela, Mar{\'\i}a Bel{\'e}n and Hern{\'a}ndez, Pilar and Orloff, Jean and
  P{\`e}ne, Olivier and Quimbay, C}.
\newblock {Standard Model CP-violation and Baryon asymmetry Part II: Finite
  Temperature}.
\newblock {\em {arXiv preprint hep-ph/9406289}}, {1994}.

\bibitem{concha_neutrino_2003}
{Gonz{\'a}lez-Garci{\'a}, M Concepci{\'o}n and Nir, Yosef}.
\newblock {Neutrino masses and mixing: evidence and implications}.
\newblock {\em {Reviews of Modern Physics}}, {75}({2}):{345}, {2003}.

\bibitem{diaz2004}
{Diaz, Rodolfo A and Martinez, R and Ochoa, F}.
\newblock {Scalar sector of the
  $\mathrm{SU}(3)_C\times\mathrm{SU}(3)_L\times\mathrm{U}(1)_X$ model}.
\newblock {\em {Physical Review D}}, {69}({9}):{095009}, {2004}.

\bibitem{diaz2005}
{Diaz, Rodolfo A and Martinez, R and Ochoa, F}.
\newblock {$\mathrm{SU}(3)_C\times\mathrm{SU}(3)_L\times\mathrm{U}(1)_X$ models
  for $\beta$ arbitrary and families with mirror fermions}.
\newblock {\em {Physical Review D}}, {72}({3}):{035018}, {2005}.

\bibitem{ecker1987GCP}
{Ecker, G. and Grimus, W. and Neufeld, H.}
\newblock {A Standard Form for Generalized (CP) Transformations}.
\newblock {\em {J. Phys.}}, {A}({20}):{L807}, {1987}.

\bibitem{branco2011}
{Branco, Gustavo Castelo and Ferreira, PM and Lavoura, L and Rebelo, MN and
  Sher, Marc and Silva, Joao P}.
\newblock {Theory and phenomenology of two-Higgs-doublet models}.
\newblock {\em {Physics reports}}, {516}({1}):{1--102}, {2012}.

\bibitem{haber2012group}
{Haber, Howard E and Surujon, Ze'ev}.
\newblock {Group-theoretic condition for spontaneous CP violation}.
\newblock {\em {Physical Review D}}, {86}({7}):{075007}, {2012}.

\bibitem{doff2006spontaneous}
{Doff, A and Pires, CA de S and da Silva, PS Rodrigues}.
\newblock {Spontaneous CP violation in the 331 model with right-handed
  neutrinos}.
\newblock {\em {Physical Review D}}, {74}({1}):{015014}, {2006}.

\bibitem{montero1999soft}
{Montero, JC and Pleitez, V and Ravinez, O}.
\newblock {Soft superweak CP violation in a 331 model}.
\newblock {\em {Physical Review D}}, {60}({7}):{076003}, {1999}.

\bibitem{weigert1997baker}
{Weigert, Stefan}.
\newblock {Baker-Campbell-Hausdorff relation for special unitary groups}.
\newblock {\em {Journal of Physics A: Mathematical and General}},
  {30}({24}):{8739}, {1997}.

\bibitem{curtright2015elementary}
{Curtright, Thomas L and Zachos, Cosmas K}.
\newblock {Elementary results for the fundamental representation of $SU(3)$}.
\newblock {\em {Reports on Mathematical Physics}}, {76}({3}):{401--404},
  {2015}.

\bibitem{Branco1999}
G.~C. Branco, L.~Lavoura, and Joao~P. Silva.
\newblock {\em {CP violation}}.
\newblock {Oxford University Press}, 1999.

\bibitem{ginzburg2005}
{Ginzburg, Ilya F and Krawczyk, Maria}.
\newblock {Symmetries of two Higgs doublet model and CP violation}.
\newblock {\em {Physical Review D}}, {72}({11}):{115013}, {2005}.

\bibitem{Brod:2013cka}
Joachim Brod, Ulrich Haisch, and Jure Zupan.
\newblock {Constraints on CP-violating Higgs couplings to the third
  generation}.
\newblock {\em JHEP}, 11:180, 2013.

\bibitem{Inoue:2014nva}
Satoru Inoue, Michael~J. Ramsey-Musolf, and Yue Zhang.
\newblock {CP-violating phenomenology of flavor conserving two Higgs doublet
  models}.
\newblock {\em Phys. Rev.}, D89(11):115023, 2014.

\bibitem{Keus:2015hva}
Venus Keus, Stephen~F. King, Stefano Moretti, and Kei Yagyu.
\newblock {CP Violating Two-Higgs-Doublet Model: Constraints and LHC
  Predictions}.
\newblock {\em JHEP}, 04:048, 2016.

\bibitem{Grimus_2000}
Walter Grimus and Lu{\'{\i}}s Lavoura.
\newblock The seesaw mechanism at arbitrary order: disentangling the small
  scale from the large scale.
\newblock {\em Journal of High Energy Physics}, 2000(11):042--042, nov 2000.

\end{thebibliography}
\bibliographystyle{unsrt}

\end{document}